\documentclass[fleqn,usenatbib]{mnras}

\usepackage{newtxtext,newtxmath}

\usepackage[T1]{fontenc}

\DeclareRobustCommand{\VAN}[3]{#2}
\let\VANthebibliography\thebibliography
\def\thebibliography{\DeclareRobustCommand{\VAN}[3]{##3}\VANthebibliography}

\usepackage{graphicx}	
\usepackage{amsmath}	
\usepackage{lscape}
\usepackage{subfig}
\usepackage{makecell}
\usepackage{threeparttable}

\title[GW Constraints on Dense Matter EOS]{Baryonic dense matter in view of gravitational-wave observations}

\author[V. B. Thapa, A. Kumar, M. Sinha]{
Vivek Baruah Thapa\thanks{thapa.1@iitj.ac.in}, 
Anil Kumar\thanks{anil.1@iitj.ac.in} 
and Monika Sinha
\thanks{ms@iitj.ac.in}
\\
Indian Institute of Technology Jodhpur, Jodhpur 342037, India
}

\pubyear{2021}

\usepackage[normalem]{ulem}  

\begin{document}
\label{firstpage}
\pagerange{\pageref{firstpage}--\pageref{lastpage}}
\maketitle

\begin{abstract}
The detection of gravitational waves (GWs) from the merger of binary neutron star (NS) events (GW170817 and GW190425) and subsequent estimations of tidal deformability play a key role in constraining the behaviour of dense matter. In addition, massive NS candidates ($\sim 2 M_{\odot}$)
along with NICER mass$-$radius measurements
also set sturdy constraints on the dense matter equation of state. Strict bounds from GWs and massive NS observations constrain the theoretical models of nuclear matter comportment at large density regimes. On the other hand, model parameters providing the highly dense matter response are bounded by nuclear saturation properties. This work analyses coupling parametrizations from two classes based on covariant density functional models: non$-$linear and density$-$dependent schemes. Considering these constraints together, we study possible models and parametrization schemes 
with the feasibility of exotic degrees of freedom in dense matter
which go well with the astrophysical observations as well as the terrestrial laboratory experiments. We show that most parametrizations with non-linear schemes do not support the observations and experiments while density-dependent scheme goes well with both. Astrophysical observations are well explained if the inclusion of heavier non-strange baryons is considered as one fraction of the dense matter particle spectrum.
\end{abstract}




\section{Introduction}\label{sec:intro}
Neutron stars (NSs), one of the most compact objects in the Universe, provide a suitable environment to study matter at densities ranging from subsaturation to few times nuclear saturation densities \citep{1996cost.book.....G, Sedrakian2007PrPNP, weber2017pulsars}.
The well-known strategy to comprehend dense matter behaviour above nuclear saturation density ($n_0$) is understanding compact astrophysical objects. Constraints from observation of compact stellar objects help the understanding of dense matter above $n_0$. One important constraint is the lower limit of maximum stellar mass of compact object family obtained from several recent observations. On the high density regime, the primary constraint on equation of state (EOS) comes from the observations of massive compact stars with masses near $2 M_{\odot}$. Observations of massive compact objects constrain the EOS to be stiff at a higher density limit. In addition, the recent NICER (Neutron star Interior Composition ExploreR) observations provide the mass and radius measurements of two compact objects (PSR J$0030+0451$ and PSR J$0740+6620$), which further constrain the EOS.
  
Massive compact star observations eradicate the softer EOSs failing to support 
the observational lower bound on maximum mass stars. 
On the other hand, the recent detection of gravitational wave (GW) emissions from binary neutron star (BNS) inspiral of the event GW170817 \citep{LIGO_Virgo2017c, LIGO_Virgo2017a, LIGO_Virgo2017b} and GW190425 \citep{2020ApJ...892L...3A} by LIGO-Virgo Collaboration (LVC) imposes bounds (although weakly) on intermediate densities of dense matter EOS.
In the BNS merger scenario, the response of a compact star to a strong gravitational field exerted by its companion is related to the EOS via induced tidal deformations during the inspiral phase. Recent observations of GW from binary star coalescence directly provide the estimate of combined dimensionless tidal deformability ($\tilde{\Lambda}$) of participating stars indicating matter to be soft at a lower density regime. 

Due to insufficient knowledge and restrictions about matter properties at density regimes higher than $n_0$, it is tough and challenging to predict the behaviour of dense matter. Several studies have been done to comprehend the NS matter composition in phenomenological \citep{1972PhRvC...5..626V, 2001A&A...380..151D,2014PhRvC..89d5807B} as well as microscopic approaches \citep{1998PhRvC..58.1804A, 2015RvMP...87.1067C, 2019PhRvC.100d5803L}. Phenomenological models such as the relativistic mean-field (RMF) model \citep{1974AnPhy..83..491W, 1977NuPhA.292..413B, 1997IJMPE...6..515S} based on effective interactions between baryons via meson exchange can conveniently describe various finite nuclei properties and can be extrapolated from subsaturation to higher density regimes \citep{2016PhR...621..127L}. Within this framework, there are a good number of admissible phenomenological models, each with a good number of parameter value sets. With recent observational constraints related to compact stars it is possible to narrow down the models of dense matter EOS. 

Observations of massive NSs open the likelihood of exotic matter such as hyperons \citep{PhysRevLett.67.2414, Bonanno2012A&A, Weissenborn2012a, Colucci_PRC_2013, Oertel2015,2016PhRvC..94c5804F, Tolos2017b,  2018MNRAS.475.4347R,2018EPJA...54..133L}, $\Delta$-resonances \citep{Drago_PRC_2014, Cai_PRC_2015, Zhu_PRC_2016, Sahoo_PRC_2018, Kolomeitsev_NPA_2017, Li_PLB_2018, Li2019ApJ, Ribes_2019, Li2020PhRvD, particles3040043} and meson ($\pi,\bar{K},\rho$-mesons) condensations \citep{particles2030025, 1982ApJ...258..306H, 1999PhRvC..60b5803G, 2001PhRvC..63c5802B, 1997PhR...280....1P, 2021EPJST.tmp...32M, PhysRevD.102.123007, PhysRevD.103.063004}) in the NS interior details of which are still not completely understood. Although hyperonization in NSs may seem to be inescapable, it leads to softening of the EOS. Following similar energetic reasoning employed for nucleation of strange baryons, another interesting aspect is the onset of non-strange $\Delta$-baryons in dense NS matter. The role of $\Delta$-resonances in NS matter is still a matter of debate \citep{Li_PLB_2018,2020PhLB..80235266M}. Inclusion of $\Delta$-baryons softens the EOS at lower matter density regimes leading to higher compactness at those densities, consequently satisfying the GW170817 event observables \citep{Li2019ApJ}. Similar to other exotic degrees of freedom, the inclusion of meson condensates also softens the EOS extensively. For recent reviews regarding the composition of compact stars, the reader may refer to \citet{2017PhR...681....1Y, 2016EPJA...52...29C, Sedrakian2017_EPJWeb, Baym_2018} and \citet{2021arXiv210514050S}.

In this work, we employ the available constraints obtained so far from compact object observations to narrow down the high-density matter models.
We explore the possible parametrization models based on constraints from terrestrial experiments and astrophysical observations (viz. massive NSs, radii estimations of NSs, 
GW emitted during the inspiral phase of BNS coalescence) 
on dense matter EOS. To construct the EOS, we consider non-linear Walecka (NLW) type and density-dependent (DD) meson-baryon couplings within covariant density functional (CDF) model with exotic degrees of freedom in addition to nucleons. 
The phenomenological EOS models based on density functional theories and realistic nuclear potentials have been analysed considering matter composition to be only nucleonic \citep{PhysRevC.98.035804, PhysRevC.99.052802, hnps2989}.
Therefore, this work will explore the novel aspects of CDF model parametrizations that satisfy the recent astrophysical observable estimations considering hyperonization of dense matter to be a viable energetic argument.

The paper is organized as follows. In Section \ref{sec:observations}, we describe a few estimated parameters of compact stellar objects from observational constraints relevant to this work. The CDF model formalism (NLW and DD), its extension to heavier baryons in $\beta$-equilibrated nuclear matter, aspects of tidal deformability of NSs and coupling parameters incorporated in this work are described in Section \ref{sec:formalism}. Section \ref{sec:results} provides the results and our conclusions and future perspectives are summarized in Section \ref{sec:summary}. 

\textit{Conventions}: We implement the natural units $G=\hbar=c=1$ throughout the work.

\section{Observational constraints}\label{sec:observations}

The understanding of dense matter above $n_0$ can be reasonably improved from recent compact star observations as the constraints on lower limit of maximum mass, observed range of mass-radius of certain compact stars and most importantly the compactness from observed tidal deformibility. The soft EOS of highly dense matter can be ruled out from the observations of massive stars viz. PSR J$1614-2230$ ($M=1.908\pm 0.016~M_\odot$) \citep{2010Natur.467.1081D, Arzoumanian_2018}, PSR J$0348+0432$ ($2.01\pm 0.04~M_\odot$) \citep{2013Sci...340..448A}, millisecond pulsar J$0740+6620$ ($2.14^{+0.20}_{-0.18}~M_\odot$ with 95$\%$ credibility \citep{2020NatAs...4...72C}, $2.08^{+0.07}_{-0.07}$ $M_\odot$ with 68.3\% credibility \citep{2021arXiv210400880F}) and PSR J$1810+1744$ ($2.13 \pm 0.04~M_\odot$ with 68$\%$ credibility) \citep{2021ApJ...908L..46R}.

GW observations also can constrain well the models of highly dense matter as already stated. For low-spin prior systems from GW170817 event, it is estimated that the combined dimensionless tidal deformability ($\tilde{\Lambda}$) parameter value has an upper and lower bounds of $900$ (TaylorF2 model) \citep{LIGO_Virgo2017c} and $400$ (AT2017gfo event) \citep{Radice2018} respectively. Reanalysis of the GW170817 data by LVC has set new limits as $110 \leq \tilde{\Lambda} \leq 720$ (PhenomPNRT model) \citep{PhysRevX.9.011001}. Another estimation of $\tilde{\Lambda}$ based on the viability of chiral effective field theory results provides the limit to be in the range $80 \leq \tilde{\Lambda} \leq 570$ \citep{Tews2018}. In addition, an ameliorated analysis of GW170817 event data implementing identical EOS for both the compact stars producing rational waveforms provides a limit on the dimensionless tidal deformability ($\Lambda$) for a $1.4M_{\odot}$ NS to be in the range $70 \leq \Lambda_{1.4} \leq 580$ with 90$\%$ credibility \citep{LIGO_Virgo2018a}. \citet{Raithel_2018} reported that the $\tilde{\Lambda} \leq 800$ constraint implies the radius of primary compact object to be $< 13$ km.

Another GW observation (GW190814) by the LIGO Livingston detector (LVC) inferred to be from a coalescence of a black-hole (BH) and lighter compact object appendage with mass of the latter to be $2.59^{+0.08}_{-0.09}~M_{\odot}$ \citep{2020ApJ...896L..44A} which falls in the `mass-gap'. The nature of lighter companion is still not resolved \citep{2020MNRAS.499L..82M, Tews_2021, 2020PhRvD.102d1301S, LI2020135812, 2021PhRvC.103b5808D, 2020arXiv201001509B, 2020PhRvC.102f5805F}.

Along with this, we get good information about mass-radius relation of compact stars from NICER observations. This space mission recently provided adequate information to estimate the mass-radius of PSR J$0030+0451$ to be in the range of $1.44^{+0.15}_{-0.14}$ $M_\odot$, $13.02^{+1.24}_{-1.06}$ km \citep{2019ApJ...887L..24M} and $1.34^{+0.15}_{-0.16}$ $M_\odot$, $12.71^{+1.14}_{-1.19}$ km (with 68.3$\%$ credibility) \citep{2019ApJ...887L..21R} respectively. 
Latest estimate of mass-radius of PSR J$0740+6620$ by NICER is in the range of $2.072^{+0.067}_{-0.066}$ $M_\odot$, $12.39^{+1.30}_{-0.98}$ km \citep{2021arXiv210506980R} and $2.08\pm0.07$ $M_\odot$, $13.71^{+2.62}_{-1.50}$ km (with 68$\%$ credibility) \citep{2021arXiv210506979M} respectively.
Several works \citep{2021arXiv210511031Z, 2021arXiv210502886B, 2021arXiv210508688P, 2021arXiv210506981R, 2021arXiv210413612S, 2021arXiv210504629L} have been performed based on different analyses on astrophysical observations inclusive of the latest NICER measurements to extract new information regarding dense matter EOS.
 
An estimation for tidal deformability from NICER observations of PSR J$0030+0451$ jointly with GW170817 event provides $240 \leq \Lambda_{1.4} \leq 730$ with radius range as $R_{1.4}=12.1^{+1.2}_{-0.8}$ km \citep{Jiang_2020}. Another recent analysis of the same PSR J$0030+0451$ data (NICER) reveals $R_{1.4}=12.32^{+1.09}_{-1.47}$ km \citep{2020PhRvD.101l3007L}. The radius constraint on the $1.4 M_{\odot}$ NSs from the GW170817 event has been derived to be in the range $10.5 \leq R_{1.4}/\text{km} \leq 13.4$ \citep{PhysRevX.9.011001} while \citet{PhysRevC.98.035804} provide the radius limit to be $11.82 \leq R_{1.4}/\text{km} \leq 13.72$. Considering similar low-spin prior systems for the GW190425 event, an upper bound of $600$ 
(PhenomPv2NRT model) has been placed on $\tilde{\Lambda}$ and the radius upper limit is derived as $R<15$ km \citep{2020ApJ...892L...3A}.  

From the analysis of GW170817 event data, bounds on matter pressure ranges are derived at $2n_0$ and $6n_0$ to be $3.5^{+2.7}_{-1.7} \times 10^{34}$ dyn-cm$^{-2}$ and $9.0^{+7.9}_{-2.6} \times 10^{35}$ dyn-cm$^{-2}$ respectively \citep{LIGO_Virgo2018a}. 
Recent results obtained from GW170817 event by Bayesian analysis suggest matter pressure at $2n_0$ to be $\sim 3.81^{+1.18}_{-2.32} \times 10^{34}$ dyn/cm$^2$ and $133 \leq \Lambda_{1.4} \leq 686$ \citep{2021EPJA...57...31L}.
The analysis of GW190425 event data reveals core matter density of the primary component involved to be in the range $3\leq n/n_0 \leq 6$ and matter pressure to be in $10^{35} \leq P(3-6~n_0) \leq 8 \times 10^{35}$ dyn/cm$^2$ range \citep{2020ApJ...892L...3A}.

We take into consideration the source properties at $90\%$ credible intervals with low-spin posterior distributions from PhenomPNRT \citep{PhysRevX.9.011001} and PhenomPv2NRT \citep{2020ApJ...892L...3A} waveform models for GW170817 and GW190425, respectively, to evaluate various GW observables.
Based on recent observable estimations of GW events, in addition to setting radius bounds on the 
NSs involved in GW170817 and GW190425 events,
we also find limits on compactness parameter for a $1.4~M_{\odot}$ NS.

\section{Formalism}\label{sec:formalism}

\subsection{EOS model}

In this section, we briefly discuss the NLW and DD CDF models implemented to contrive the EOS in this work. The dense matter possible constituents considered here are nucleons ($N\equiv n,p$), hyperons ($Y\equiv \Lambda^0, \Sigma^{\pm,0},\Xi^{-,0}$) and $\Delta$-resonances ($\Delta \equiv \Delta^{++}, \Delta^{+}, \Delta^{-}, \Delta^{0}$) alongside leptons ($l\equiv e^-,\mu^{-}$) to maintain $\beta$-equilibrium. The interaction between non-strange baryons are described via the exchange of isoscalar-scalar $\sigma$, isoscalar-vector $\omega$, and isovector-vector $\rho$ mesons. For the hyperonic sector interactions, an additional hidden strangeness isoscalar-vector $\phi$ meson is taken into consideration. In general, the total Lagrangian density describing the baryon-meson interactions is given by \citep{1999PhRvC..60b5803G,2001PhRvC..63c5802B,Li_PLB_2018}
\begin{equation}\label{eqn.1}
\begin{aligned}
\mathcal{L} & = \sum_{b\equiv N,Y} \bar{\psi}_b(i\gamma_{\mu} D^{\mu}_{(b)} - m^{*}_b) \psi_b + \sum_{l} \bar{\psi}_l (i\gamma_{\mu} \partial^{\mu} - m_l)\psi_l 
\\
& + \sum_{\Delta} \bar{\psi}_{\Delta \nu}(i\gamma_{\mu} D^{\mu}_{(\Delta)} - m^{*}_{\Delta}) \psi^{\nu}_{\Delta} + \frac{1}{2}(\partial_{\mu}\sigma\partial^{\mu}\sigma - m_{\sigma}^2 \sigma^2) \\
 & -  \frac{1}{4}\omega_{\mu\nu}\omega^{\mu\nu} + \frac{1}{2}m_{\omega}^2\omega_{\mu}\omega^{\mu} - \frac{1}{4}\boldsymbol{\rho}_{\mu\nu} \cdot \boldsymbol{\rho}^{\mu\nu} + \frac{1}{2}m_{\rho}^2\boldsymbol{\rho}_{\mu} \cdot \boldsymbol{\rho}^{\mu}  \\
 & - \frac{1}{4}\phi_{\mu\nu}\phi^{\mu\nu} + \frac{1}{2}m_{\phi}^2\phi_{\mu}\phi^{\mu} - \text{U}(\sigma)
\end{aligned}
\end{equation}
where, the last term $\text{U}(\sigma)$ corresponds to the self-interactions of scalar mesons only for NLW CDF models. $\psi_b,~\psi_{l},~\psi_{\Delta}^{\nu}$ represent Dirac-fields of the baryon-octet, leptons and Schwinger-Rarita fields of $\Delta$-quartet respectively. $m_b,~m_l$, and $m_{\Delta}$ denote the bare masses of baryon octet, leptons and $\Delta$-quartet species respectively. The covariant derivative mentioned in eq.-\eqref{eqn.1} is given by
\begin{equation}\label{eqn.2}
D_{\mu (j)} = \partial_\mu + ig_{\omega j} \omega_\mu + ig_{\rho j} \boldsymbol{\tau}_{j3} \cdot \boldsymbol{\rho}_{\mu} + ig_{\phi j} \phi_\mu
\end{equation}
with $j$ denoting the baryon octet ($b$) and $\Delta$-quartet ($\Delta$). The coupling constants are represented by $g_{pj}$ with index-$p$ being the considered mesons. $\boldsymbol{\tau}_{j3}$ is the iso-spin projection of third component of isovector-vector meson fields. The scalar self-interaction term for NLW CDF model is given by
\begin{equation}\label{eqn.3}
\text{U}(\sigma) = \frac{1}{3} g_2 \sigma^3 + \frac{1}{4} g_3 \sigma^4
\end{equation}
where $g_2$ and $g_3$ are the coefficients of self-interactions. 

With the monotonic increase in baryon chemical potentials interior to NSs, the nucleonic matter may well transform to heavier strange and non-strange baryons leading to new hadronic degrees of freedom. 
The conditions which are necessary to maintain strong $\beta$-equilibrium between different particle species without strangeness being conserved are \citep{2001PhRvC..63c5802B, Drago_PRC_2014}
\begin{equation}\label{eqn.7}
\begin{aligned}
\mu_e & = \mu_n - \mu_p = \mu_{\mu}, \quad \mu_{\Sigma^+} = \mu_{\Delta^+} = \mu_p, \\
\mu_{\Sigma^-} & = \mu_{\Xi^-} = \mu_{\Delta^-} = \mu_n + \mu_e, \quad \mu_{\Delta^{++}} = \mu_p - \mu_e, \\
\mu_{\Sigma^0} & =\mu_{\Xi^0}=\mu_{\Lambda^0} = \mu_{\Delta^0}= \mu_n
\end{aligned}
\end{equation}
with $\mu_j$ denoting chemical potential of the $j$-th baryon and defined as
\begin{equation}\label{eqn.8}
\begin{aligned}
& \mu_{j} = \sqrt{p_{F_j}^2 + m_{j}^{*2}} + \Sigma_{B}.
\end{aligned}
\end{equation}
Here, $\Sigma_B = \Sigma^{0} + \Sigma^{r}$ represents the vector self energies with
$\Sigma^{0} = g_{\omega j}\omega_{0} + g_{\phi j}\phi_{0} + g_{\rho j} \boldsymbol{\tau}_{j3} \rho_{03}$ and
$\Sigma^{r}$ denotes the rearrangement term (present in DD CDF models only) necessary to maintain thermodynamic consistency which is given by
\begin{equation}
\begin{aligned}\label{eqn.rear}
	\Sigma^{r} & = \sum_{b} \left[ \frac{\partial g_{\omega b}}{\partial n}\omega_{0}n_{b} - \frac{\partial g_{\sigma b}}{\partial n} \sigma n_{b}^s + \frac{\partial g_{\rho b}}{\partial n} \rho_{03} \boldsymbol{\tau}_{b3} n_{b} \right. \\
	& \left. + \frac{\partial g_{\phi b}}{\partial n}\phi_{0}n_{b} \right] + \sum_{\Delta} (\psi_b \longrightarrow \psi_{\Delta}^{\nu}),
\end{aligned}
\end{equation}
with $n^s= \langle\bar{\psi} \psi \rangle$ and $n=\langle\bar{\psi} \gamma^0 \psi \rangle$ denoting the scalar and vector (number) densities respectively.

Two additional constraints, viz. charge neutrality and global baryon number conservation, are administered while evaluating the EOS self-consistently. The re-arrangement term considered in DD CDF model contributes explicitly to the pressure term. For details, the readers may refer to \cite{2001PhRvC..64b5804H}.

\subsection{Tidal deformability}

The compact stars in a binary system experience tidal deformations due to the gravitational fields of their respective companions. These tidal effects can be quantified in terms of tidal deformability parameter ($\lambda$) defined as the ratio of induced mass quadrupole moment $Q_{ij}$ to external perturbing tidal field $\mathcal{E}_{ij}$ \citep{Hinderer_2008, PhysRevD.77.021502, PhysRevD.81.123016}, 
\begin{equation}\label{eqn.9}
\lambda = - \frac{Q_{ij}}{\mathcal{E}_{ij}} = \frac{2}{3}k_2 R^5,
\end{equation}
where,
\begin{equation}\label{eqn.10}
\begin{aligned}
k_2 = & \frac{8 C^5}{5}(1-2C^2)[2+2C(y-1)-y] \cdot \{2C [6-3y+3C(5y-8)] \\
& + 4C^3 [13 - 11y + C(3y-2) + 2C^2(1+y)] + 3(1-2C^2) \cdot \\
& [2 - y + 2C(y-1)] \log (1-2C)\}^{-1}
\end{aligned}
\end{equation}
with $C=M/R$ being the compactness parameter, $M$ and $R$ being mass and radius of the star respectively and $k_2$, the EOS dependent tidal Love number. $y=y(R)$ is the function obtained after solving the differential equation \citep{PhysRevD.80.084018, PhysRevD.81.084016, PhysRevC.95.015801},
\begin{equation}\label{eqn.11}
r \frac{dy(r)}{dr} + y(r)^2 + y(r)F(r) + r^2 Q(r) = 0,
\end{equation}
where the functions are
\begin{equation}\label{eqn.12}
F(r) = \frac{r - 4\pi r^3 [\varepsilon(r) - P(r)]}{r - 2M(r)},
\end{equation}
\begin{equation}\label{eqn.13}
\begin{aligned}
Q(r) = & \frac{4\pi r [5\varepsilon(r) + 9P(r) + \frac{\varepsilon(r) + P(r)}{\partial P(r)/\partial \varepsilon(r)}]}{r - 2M(r)} - 4 \left[ \frac{M(r) + 4\pi r^3 P(r)}{r^2 (1 - 2M(r)/r)} \right].
\end{aligned}
\end{equation}
\begin{table*} 
\caption{The nuclear properties of the CDF models at respective $n_0$.}
\centering
\begin{tabular}{cccccccc}
\hline \hline
\multicolumn{2}{c}{CDF Model} & $n_0$ & $-E_0$ & $K_0$ & $E_{\text{sym}}$ & $L_{\text{sym}}$ & $K_{\text{sym}}$ \\
& & (fm$^{-3}$) & (MeV) & (MeV) & (MeV) & (MeV) & (MeV) \\
\hline
& GM1 & 0.153 & $16.30$ & 300.00 & 32.50 & 93.857 & 17.91 \\
& GM2 & 0.153 & $16.30$ & 300.00 & 32.50 & 89.289 & $-$ 11.98 \\
& GM3 & 0.153 & $16.30$ & 240.00 & 32.50 & 89.627 & $-$ 6.46 \\
& NL3 & 0.148 & $16.29$ & 271.76 & 37.40 & 118.317 & 100.53 \\
NLW & NL3-II & 0.149 & $16.28$ & 272.15 & 37.70 & 119.563 & 103.19 \\
& NL-SH & 0.146 & $16.346$ & 355.36 & 36.10 & 113.654 & 79.81 \\
& NL-RA1 & 0.1466 & $16.15$ & 285.00 & 36.10 & 115.305 & 95.57 \\
& NL3* & 0.150 & $16.31$ & 258.27 & 38.68 & 122.71 & 105.73 \\
& GMT & 0.145 & $16.30$ & 281.00 & 36.90 & 112.796 & 63.04 \\
\hline
& DD1 & 0.1487 & $16.021$ & 240.00 & 31.60 & 55.949 & $-$ 95.24 \\
& DD2 & 0.149065 & $16.02$ & 242.70 & 32.73 & 54.966 & $-$ 93.24 \\
& DD-ME1 & 0.152 & $16.20$ & 244.50 & 33.10 & 55.370 & $-$ 101.07 \\
& DD-ME2 & 0.152 & $16.14$ & 250.89 & 32.30 & 51.253 & $-$ 87.31 \\
DD & PKDD & 0.149552 & $16.267$ & 262.181 & 36.79 & 90.139 & $-$ 80.56 \\
& TW99 & 0.153 & $16.247$ & 240.00 & 33.39 & 55.309 & $-$ 124.68 \\
& DDV & 0.151 & $16.097$ & 240.00 & 33.589 & 71.463 & $-$ 93.97 \\
& DDF & 0.1469 & $16.024$ & 223.10 & 31.60 & 55.919 & $-$ 139.66 \\
& DD-MEX & 0.152 & $16.14$ & 267.059 & 32.269 & 49.576 & $-$ 71.47 \\
\hline
\end{tabular}
\label{tab:1}
\end{table*} 
Due to paramount dependence on the stellar radius, $\lambda$ imposes stringent constraints on dense matter EOS. $\lambda$ or, equivalently $k_2$ provides the ease of induced deformation estimate of bulk matter. This parameter is evaluated self-consistently alongside the Tolman-Oppenheimer-Volkov (TOV) equations \citep{1996cost.book.....G}. Another dimensionless quantity $\Lambda$ is much more expedient as it relates $\lambda$ with $C$ through the relation 
\begin{equation}\label{eqn.14}
\Lambda=\frac{\lambda}{M^5}=\frac{2}{3} \frac{k_2}{C^5}.  
\end{equation}

GW signal encodes information regarding deformation of both compact objects in the binary system as the weighted tidal deformability ($\tilde{\lambda}$) and is given by
\citep{Hinderer_2008, PhysRevD.81.123016}
\begin{equation}\label{eqn.15}
\tilde{\lambda} = \frac{1}{26} \left[ \frac{M_1 + 12 M_2}{M_1} \lambda_1 + \frac{M_2 + 12 M_1}{M_2} \lambda_2 \right]
\end{equation}
where $\lambda_1$ and $\lambda_2$ are the tidal deformabilities corresponding to stars with masses $M_1$ and $M_2$ respectively. In order to relate to $\Lambda$, we incorporate the combined dimensionless tidal deformability ($\tilde{\Lambda}$) defined as \citep{PhysRevLett.112.101101}
\begin{equation}\label{eqn.16}
\begin{aligned}
\tilde{\Lambda} & = 32 \frac{\tilde{\lambda}}{(M_1 + M_2)^5} \\
& = \frac{16}{13} \frac{(M_1 + 12 M_2) M_{1}^4 \Lambda_1 + (M_2 + 12 M_1) M_{2}^4 \Lambda_2}{(M_1 + M_2)^5} .
\end{aligned}
\end{equation}

\subsection{Coupling parameters} \label{cp}

In this work, we adopt the NLW model with GM1, GM2, GM3 \citep{PhysRevLett.67.2414}, NL3, NL3-II \citep{1997PhRvC..55..540L}, NL-SH \citep{SHARMA1993377}, NL-RA1 \citep{PhysRevC.63.044303}, NL3* \citep{LALAZISSIS200936}, GMT \citep{2000NuPhA.674..553P} parametrizations and the DD CDF model with DD1 \citep{PhysRevC.71.064301}, DD2 \citep{PhysRevC.81.015803}, DD-ME1 \citep{PhysRevC.66.024306}, DD-ME2 \citep{2005PhRvC..71b4312L}, PKDD \citep{2004PhRvC..69c4319L}, TW99 \citep{TYPEL1999331}, DDV \citep{Typel:2020ozc}, DDF \citep{PhysRevC.74.035802}, DD-MEX \citep{TANINAH2020135065} parametrizations for meson-baryon couplings. 

\begin{figure} 
  \begin{center}
\includegraphics[width=8.5cm,keepaspectratio ]{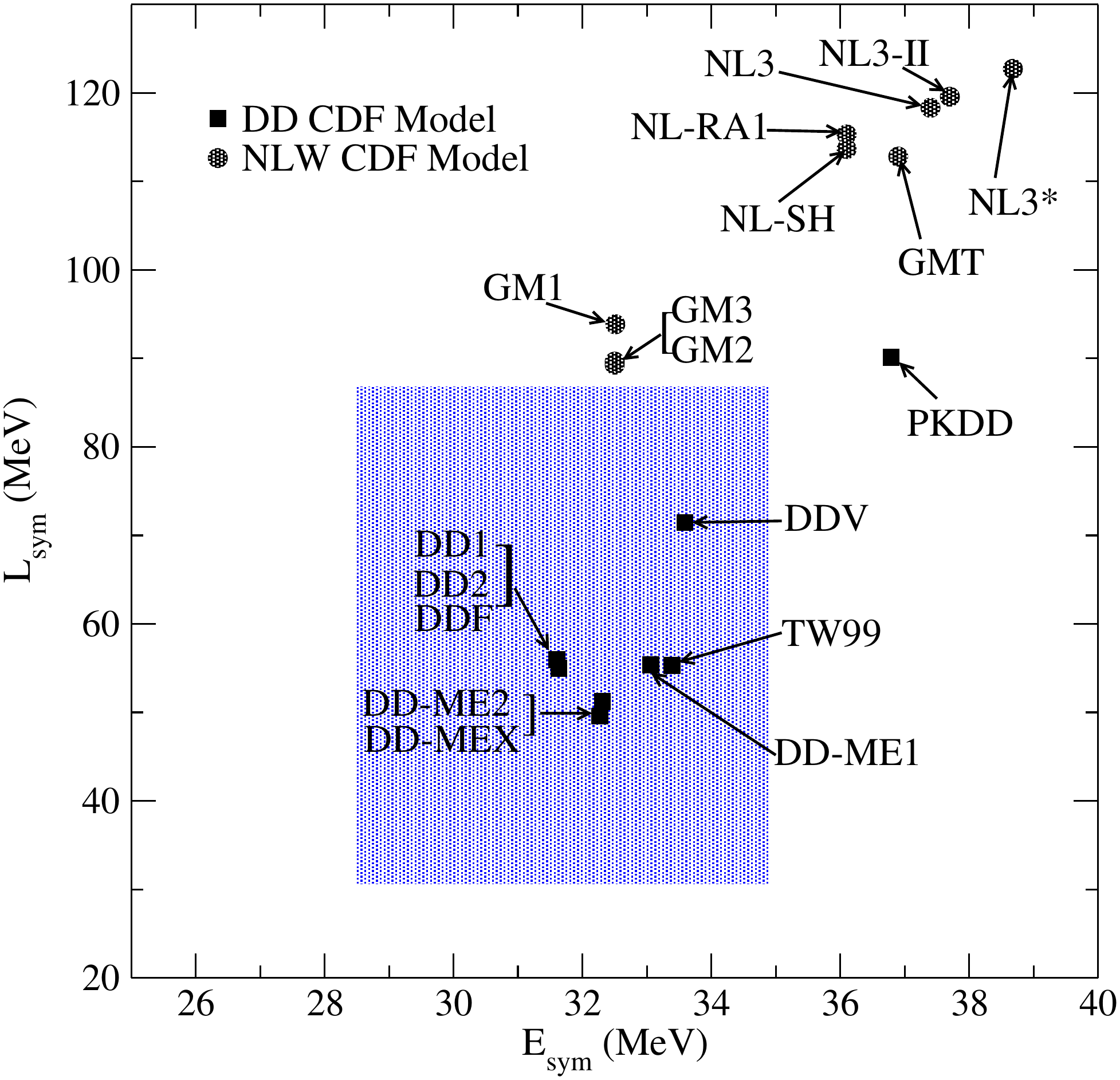}
\caption{Symmetry energy coefficient ($E_{\text{sym}}$) and its corresponding slope parameter ($L_{\text{sym}}$) for all the parametrization models considered in this work. The shaded region represents the current empirical range in $E_{\text{sym}}-L_{\text{sym}}$ plane following recent experimental and microscopic model calculations \citep{2017RvMP...89a5007O}.}
\label{fig-012}
\end{center}
\end{figure} 

Comparison of experimental data \citep{2017RvMP...89a5007O} from finite nuclei and heavy-ion collisions with different microscopic model calculations have provided bounds on nuclear saturation properties of symmetric nuclear matter (SNM):
\begin{enumerate}
\item Incompressibility, $ 210$ MeV $\leq K_0(n_0) \leq 280$ MeV,
\item Symmetry energy coefficient, $28.5$ MeV $\leq E_{\text{sym}} (n_0) \leq 34.9$ MeV,
\item Slope parameter of $E_{\text{sym}}$, $30.6$ MeV $\leq L_{\text{sym}} (n_0) \leq 86.8$ MeV.
\end{enumerate}
Table-\ref{tab:1} displays the nuclear saturation properties obtained for models with different CDF parametrizations considered in this work with $E_0$, the saturation energy.
The bounds on curvature of symmetry energy, $K_{\text{sym}}(n_0)=-111.8 \pm 71.3$ MeV \citep{2017PhRvC..96b1302M}, $-85^{+82}_{-70}$ MeV \citep{2019ApJ...887...48B} and $-102^{+71}_{-72}$ MeV \citep{2020arXiv200203210Z} deduced from nuclear data and various NS observations provide additional constraint on dense matter EOS.

\begin{figure*}%
    \centering
    \subfloat{\includegraphics[width=8.5cm,keepaspectratio ]{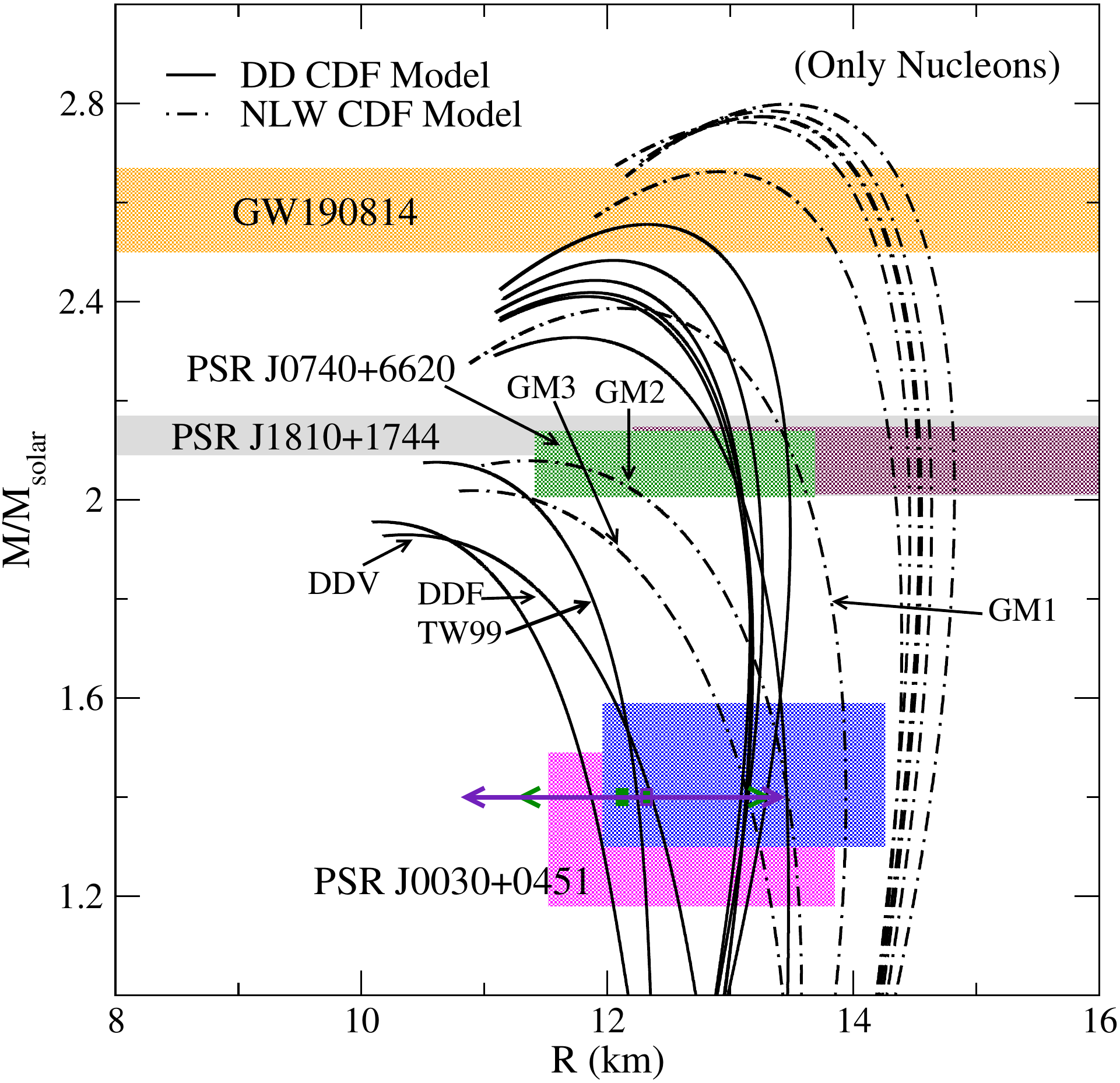}}%
    \qquad
    \subfloat{\includegraphics[width=8.5cm,keepaspectratio ]{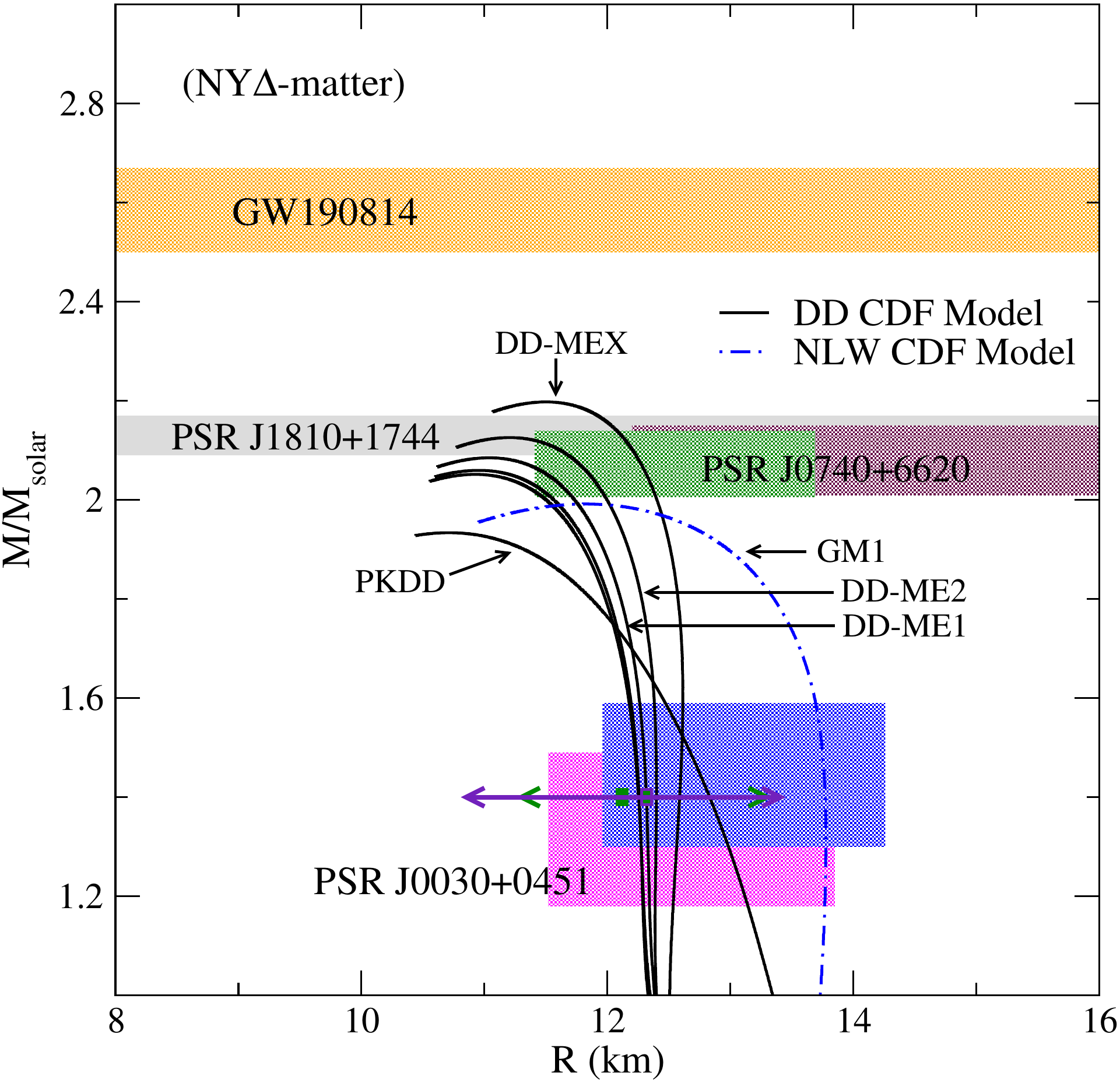}}%
    \caption{The family of solutions of TOV equations for matter composed of, left-hand panel: only nucleons and right-hand panel: $\Delta$-admixed hypernuclear matter alongside leptons to maintain $\beta$-equilibrium. The solid curves denote the $M$-$R$ curves for different density-dependent model parametrizations, while the dot-dashed curves denote the cases with non-linear model parametrizations. The astrophysical constraints from GW190814 \citep{2020ApJ...896L..44A}, PSR J$1810+1744$ \citep{2021ApJ...908L..46R}, PSR J$0030+0451$ \citep{2019ApJ...887L..24M, 2019ApJ...887L..21R, Jiang_2020, 2020PhRvD.101l3007L}, PSR J$0740+6620$ \citep{2021arXiv210506980R, 2021arXiv210506979M} are represented by the shaded regions.}%
    \label{fig-001}%
\end{figure*}

The density-dependent meson-nucleon couplings implemented in DD CDF model are defined as \citep{TYPEL1999331, PhysRevC.66.024306},
\begin{equation}\label{eqn.17}
g_{i N}(n)= g_{i N}(n_{0}) f_i(n/n_0) \quad \quad \text{for }i=\sigma,\omega
\end{equation}
where, $n$ is the baryon number density and
\begin{equation}\label{eqn.18}
f_i(n/n_0)= a_i \frac{1+b_i (n/n_0+d_i)^2}{1+c_i (n/n_0 +d_i)^2}.
\end{equation}
In the case of $\rho$-meson couplings, the functional is defined as
\begin{equation}\label{eqn.19}
g_{\rho N}(n)= g_{\rho N}(n_{0}) e^{-a_{\rho}(n/n_0 -1)}.
\end{equation}
For the coefficient values in equations \eqref{eqn.17}-\eqref{eqn.19}, the readers may refer to \citet{PhysRevC.71.064301, PhysRevC.81.015803, PhysRevC.66.024306, 2005PhRvC..71b4312L, 2004PhRvC..69c4319L, TYPEL1999331, Typel:2020ozc, PhysRevC.74.035802} and \citet{TANINAH2020135065}. These coefficients are associated with different DD CDF model parametrizations and fitted to reproduce various nuclei properties. The hidden strangeness meson $\phi$ does not couple with nucleons, so $g_{\phi N}=0$.

The meson-hyperon and meson-$\Delta$ couplings are considered similar to the density-dependence footing in the case of nucleons in DD CDF model. In the case of the meson-hyperon vector couplings, we implement the SU(6) symmetry and quark counting rule \citep{SCHAFFNER199435}. 
For the scalar meson-hyperon couplings, we consider the optical potentials of $\Lambda,\Sigma$ and $\Xi$-hyperons in SNM to be $-28,+30$ and $-14$ MeV, respectively, at nuclear saturation \citep{Feliciello_2015, RevModPhys.88.035004}. 
The optical potential depths $U_{\Xi}^{(N)} (n_0)=-18$ MeV and $U_{\Lambda}^{(N)} (n_0) =-30$ MeV \citep{2000PhRvC..62c4311S, 2015ApJ...808....8G} are also widely implemented in dense matter studies. 
Recently \citet{2021arXiv210400421F} reported an attractive optical potential depth of $\Xi$-hyperons in SNM to be $\gtrsim -20$ MeV. 
For a more recent review on the aspects of strangeness in dense matter, the reader may refer to \citet{2020PrPNP.11203770T}.

Due to scarce information regarding the $\Delta$-nucleon interactions, we treat the meson-$\Delta$ resonances couplings as parameters. \citet{PhysRevC.81.035502, KOCH1985765, WEHRBERGER1989797} have reported the data to constrain meson-$\Delta$ baryon couplings at $n_0$ based on pion-nucleus scattering, electron scattering on nuclei and excitation studies of $\Delta$-quartet experiments. Recent reviews \citep{Drago_PRC_2014, PhysRevC.74.035802, Kolomeitsev_NPA_2017} on this aspect have reported the $\Delta$-potential ($V_{\Delta}$) in nuclear medium to be in the range $-30~\text{MeV} + V_N \leq V_{\Delta} \leq V_N$ ($V_N$ being the nucleon potential), the values of factor $x_{\sigma \Delta} - x_{\omega \Delta}$ to be between $0$ and $0.2$ with $x_{\sigma \Delta}=g_{\sigma \Delta}/ g_{\sigma N}$ and $x_{\omega \Delta}=g_{\omega \Delta}/ g_{\omega N}$. Many works \citep{Drago_PRC_2014, Kolomeitsev_NPA_2017, Li_PLB_2018, Ribes_2019, PhysRevC.75.035806, Cai_PRC_2015, 2021PhLB..81436070R} have considered the ranges for $x_{\omega \Delta} \in [0.6-1.2]$ and $x_{\rho \Delta} \in [0.5-3.0]$. 
For recent development regarding $\Delta$-potential in dense matter, the reader may refer to \citet{2021arXiv210108679C}.
In the ensuing discussion, we will consider $x_{\omega \Delta}= 1.10$, $x_{\rho \Delta}=1.00$ for the vector meson-$\Delta$ baryon couplings and $x_{\sigma \Delta}=1.20$ for the scalar coupling. $\Delta$-quartet being non-strange baryons does not couple with $\phi$-meson, so $g_{\phi \Delta}=0$.

\section{Constraints on dense matter models}\label{sec:results}

Next, we sort out the different EOSs with various possible compositions and parametrizations based on terrestrial experimental and stellar observational values. From table-\ref{tab:1} it can be noticed that all the models reproduce $n_0$ and $E_0$ in the correct range of empirical values. However, among the NLW parametrizations considered in this work, GM1 and GM2 cannot reproduce the empirical range of $K_0$ at $n_0$ mentioned in Section \ref{cp}, while all the DD CDF models considered in this work satisfy the empirical range of $K_0$ at $n_0$. On the other hand, the empirical range of symmetry energy coefficient is satisfied by GM1, GM2 and GM3 parameter sets among NLW models, while in the case of DD models, all except PKDD parametrization lies within the bounds of this particular saturation property. The current empirical bound on $L_{\text{sym}}$ is satisfied by all DD parametrization models considered in this work except PKDD parametrization. 
In addition, the current empirical bounds on $K_{\text{sym}}$ are satisfied by GM2, GM3 (among NLW models) and all DD parameter sets considered in this work.
Fig.-\ref{fig-012} displays the parameter sets which are compatible with the current bounds on $E_{\text{sym}}$ and $L_{\text{sym}}$. All the DD models (except PKDD) considered in this work satisfy the current bounds on nuclear saturation properties. 

\begin{table} 
\caption{Summary of TOV results evaluated from parameter sets considered in this work (pure $N$-matter, refer to left-hand panel of fig.-\ref{fig-001}). Fulfillment of the mass-radius constraints from various astrophysical observations are marked by $+(-)$.}
\centering
\begin{threeparttable}[b]
\begin{tabular}{cccccc}
\hline \hline
\multicolumn{2}{c}{CDF Model} & \parbox[t]{1mm}{\rotatebox[origin=c]{90}{ PSR J$0030+0451$ }} & \parbox[t]{1mm}{\rotatebox[origin=c]{90}{ PSR J$0740+6620$ }} & \parbox[t]{1mm}{\rotatebox[origin=c]{90}{ PSR J$1810+1744$ }} & \thead{GW190814 \\ (secondary)\tnote{$^{**}$} } \\
\hline
& GM1 & $+$ & $+$ & $+$ & $-$ \\
& GM2 & $+$ & $+$ & $-$ & $-$ \\
& GM3 & $+$ & $-$ & $-$ & $-$ \\
& NL3 & $-$ & $+$ & $+$ & $+$ \\
NLW & NL3-II & $-$ & $+$ & $+$ & $+$ \\
& NL-SH & $-$ & $+$ & $+$ & $+$ \\
& NL-RA1 & $-$ & $+$ & $+$ & $+$ \\
& NL3* & $-$ & $+$ & $+$ & $+$ \\
& GMT & $-$ & $+$ & $+$ & $+$ \\
\hline
& DD1 & $+$ & $+$ & $+$ & $-$ \\
& DD2 & $+$ & $+$ & $+$ & $-$ \\
& DD-ME1 & $+$ & $+$ & $+$ & $-$ \\
& DD-ME2 & $+$ & $+$ & $+$ & $-$ \\
DD & PKDD & $+$ & $+$ & $+$ & $-$ \\
& TW99 & $+$ & $-$ & $-$ & $-$ \\
& DDV & $+$ & $-$ & $-$ & $-$ \\
& DDF & $+$ & $-$ & $-$ & $-$ \\
& DD-MEX & $+$ & $+$ & $+$ & $+$ \\
\hline
\end{tabular}
  \begin{tablenotes}
    \item[$^{**}$] The nature of secondary component of GW190814 is still in tension
  \end{tablenotes}
  \end{threeparttable}
\label{tab:4}
\end{table} 
The mass-radius ($M$-$R$) relations corresponding to the various parametrizations for only nucleonic matter composition, are shown in left panel of fig.-\ref{fig-001} obtained by solving the TOV 
equations for spherically symmetric, non-rotating stars. 
For the crustal region, we consider the \citet{1971ApJ...170..299B} EOS. The transition from crust to the core is modelled in a way that is thermodynamically consistent following \citet{2016PhRvC..94c5804F}.
We consider the recent massive NS observation (PSR J$1810+1744$) as the lower bound for maximum mass configurations. If we consider pure nucleonic matter, all the EOSs considered in this work except a few satisfy the lower bound constraint of maximum mass as evident from the left-hand panel of fig.-\ref{fig-001}. 
GM2 ($M_{\text{max}}\sim2.08~M_{\odot}$), GM3 ($2.02~M_{\odot}$) among NLW type and TW99 ($2.07~M_{\odot}$), DDV ($1.93~M_{\odot}$), DDF ($1.96~M_{\odot}$) among DD type models fail to satisfy this mass constraint. However, the constraints obtained from NICER observations (PSR J$0030+0451$) are seen to be not satisfied by the NLW models satisfying mass constraint, except the GM1 parametrization, while all the DD type models satisfying the mass constraint satisfy this constraint too. Moreover, these models satisfy 
the recent NICER results of PSR J$0740+6620$ simultaneously. Here, PSR J$0740+6620$ observation suggests the dense matter at higher densities is repulsive enough to produce large radii of heavier NSs. Most of the NLW models not satisfying the NICER observations of PSR J$0030+0451$ are observed to satisfy this recent constraint with a wider radii range. However, 
even though they are observed to fulfil the constraint from \citet{2021arXiv210506979M}, they fail to fulfil another NICER measurement of PSR J$0030+0451$ from \citet{2019ApJ...887L..24M, 2019ApJ...887L..21R, Jiang_2020, 2020PhRvD.101l3007L}. 
Stiffer EOSs ($N$-matter) obtained from NLW type parametrizations (except GM1, GM2, GM3) and DD-MEX (DD type) yield $M_{\text{max}} \geq 2.5~M_{\odot}$ satisfying the GW190814 event secondary component's mass constraint. Although this constraint is not unequivocal as the nature of secondary compact object is not found to be NS explicitly \citep{2020PhRvD.102d1301S, LI2020135812}.
The TOV results involving different parametrizations with pure $N$-matter are summarized in table-\ref{tab:4}.
As a result of EOS softening due to the inclusion of hyperons and $\Delta$-quartet, the only NLW model EOS with the GM1 parametrization, satisfying all mass-radius constraints, produce stars that fail to fulfil the lower bound constraint of maximum mass as shown in the right-hand panel of fig.-\ref{fig-001}. Except PKDD ($M_{\text{max}}\sim1.93~M_{\odot}$), DD1 ($2.05~M_{\odot}$) and DD2 ($2.06~M_{\odot}$) parametrizations all other parametrizations from DD CDF models which satisfy the joint constraints from PSR J$1810+1744$, PSR J$0740+6620$ and PSR J$0030+0451$ with pure $N$-matter satisfy the mass constraints even after softening due to hyperonization. The incorporation of $\Delta$-quartet reduces the radii of NSs, enhancing their respective compactness parameter. 

\begin{figure} 
  \begin{center}
\includegraphics[width=8.5cm,keepaspectratio ]{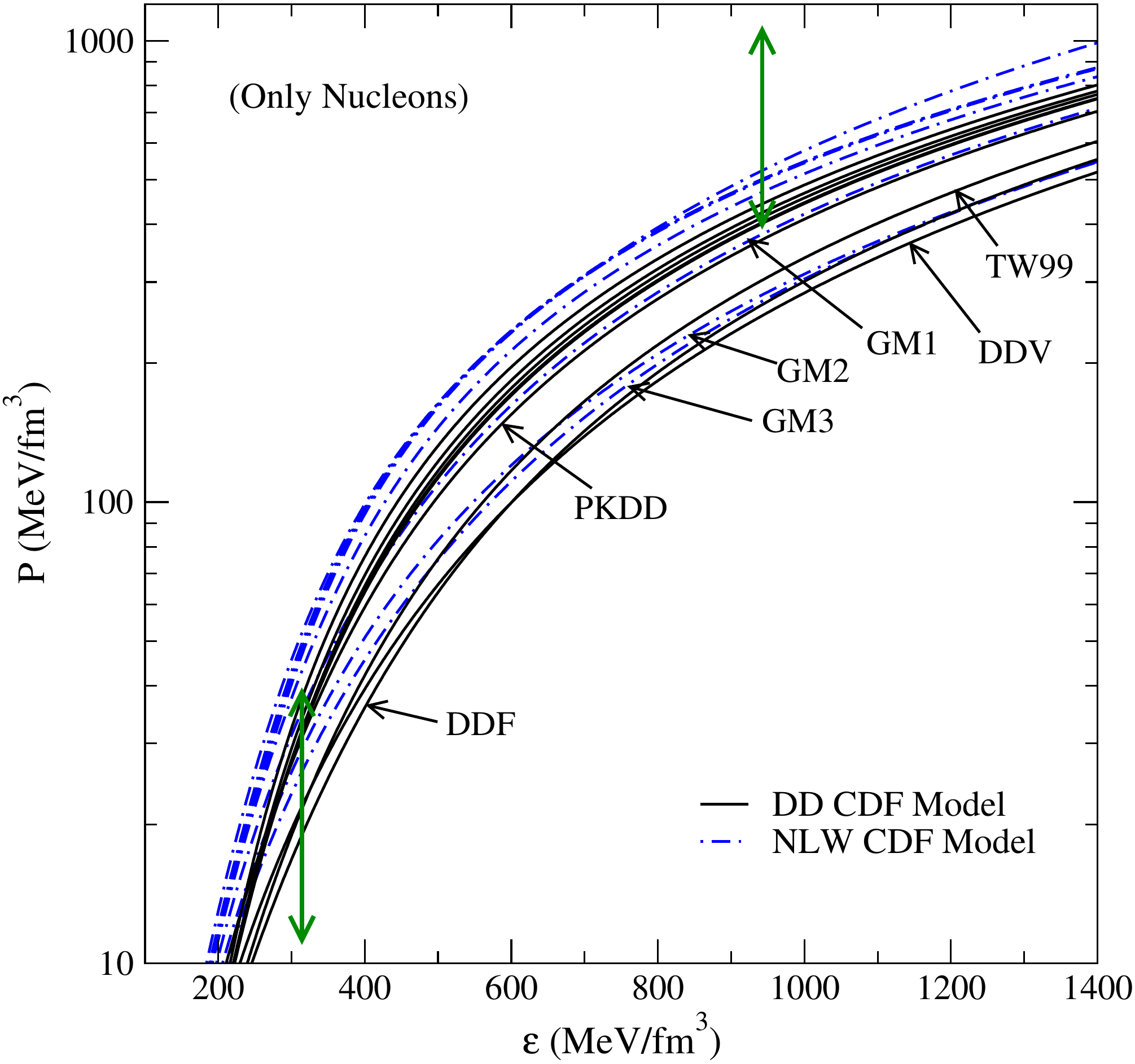}
\caption{The variation of matter pressure as a function of energy density (EOS) for $\beta$-equilibrated matter composed of only nucleons and leptons. The solid curves denote the EOSs for different density-dependent model parametrizations, while the dot-dashed curves denote the cases with non-linear model parametrizations. The constraints on matter pressure evaluated from GW170817 event are denoted by the vertical lines and provided in sec.-\ref{sec:intro}.}
\label{fig-003}
\end{center}
\end{figure}

\begin{figure*} 
  \begin{center}
\includegraphics[width=12.2cm,keepaspectratio ]{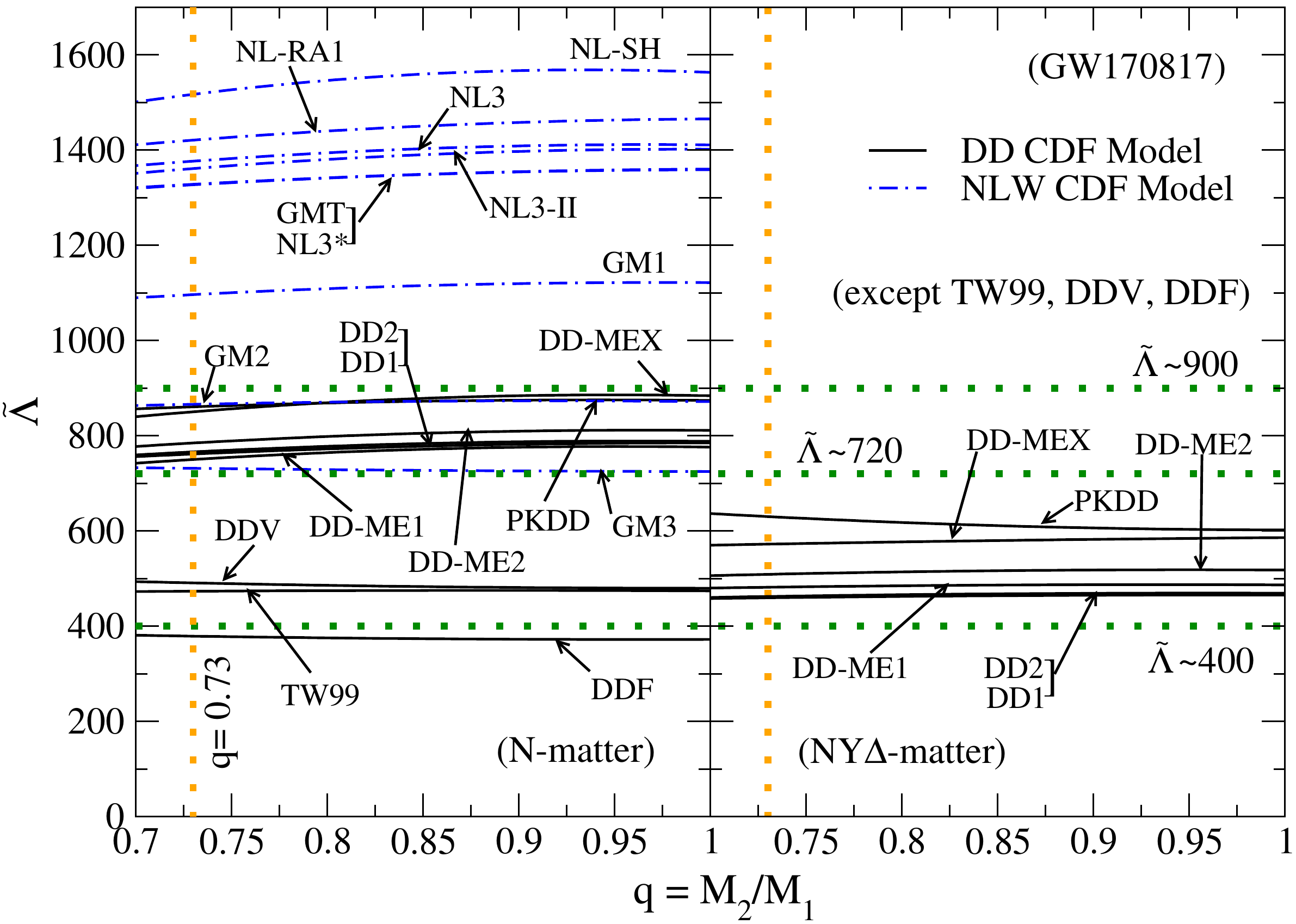}
\caption{Combined dimensionless tidal deformability as a function of mass-ratio ($q$) considering a fixed chirp mass, $\mathcal{M}=1.188~M_{\odot}$ (GW170817 event). Left panel: nucleonic matter, right panel: $\Delta$-resonances and baryon octet matter. The solid, dot-dashed lines depicts the DD and NLW type parametrizations respectively. The parametrizations yielding soft EOSs are not shown in the right panel. The horizontal dotted lines denote bounds on $\tilde{\Lambda}$ (with $90\%$ credibility, low-spin priors). The vertical dotted lines represent the mass-ratio, $q=0.73$ boundary \citep{PhysRevX.9.011001}.}
\label{fig-004}
\end{center}
\end{figure*}

\begin{figure} 
  \begin{center}
\includegraphics[width=8.5cm,keepaspectratio ]{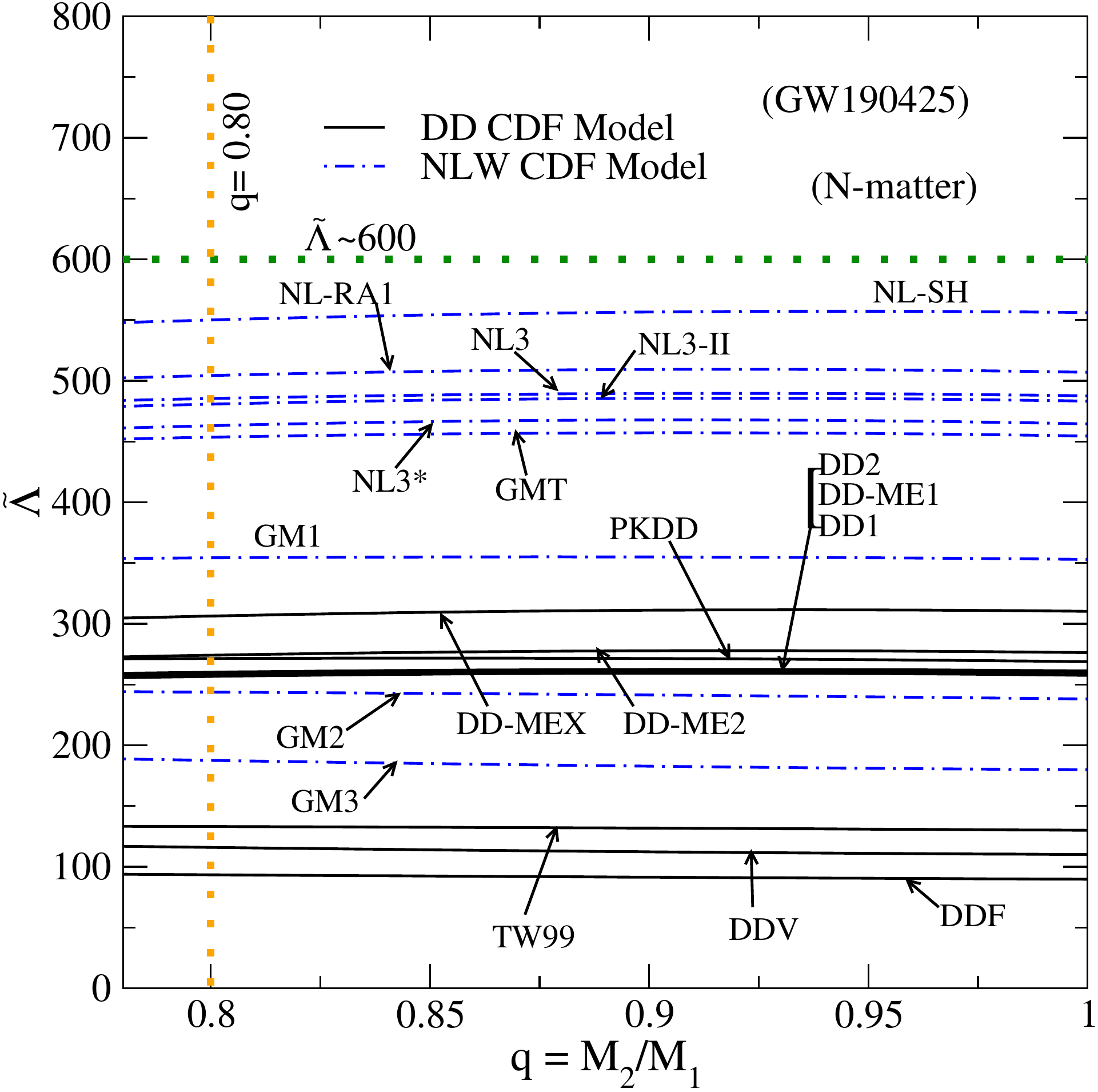}
\caption{Similar to fig.-\ref{fig-004} but considering a fixed chirp mass, $\mathcal{M}=1.43~M_{\odot}$ (GW190425 event) with pure nucleonic matter. The solid, dot-dashed lines depicts the DD and NLW type parametrizations respectively. $\tilde{\Lambda}$ constraints are similar to fig.-\ref{fig-004}. The vertical dotted lines represent the mass-ratio, $q=0.80$ boundary \citep{2020ApJ...892L...3A}.}
\label{fig-005}
\end{center}
\end{figure}

\begin{figure*} 
  \begin{center}
\includegraphics[width=12.20cm,keepaspectratio ]{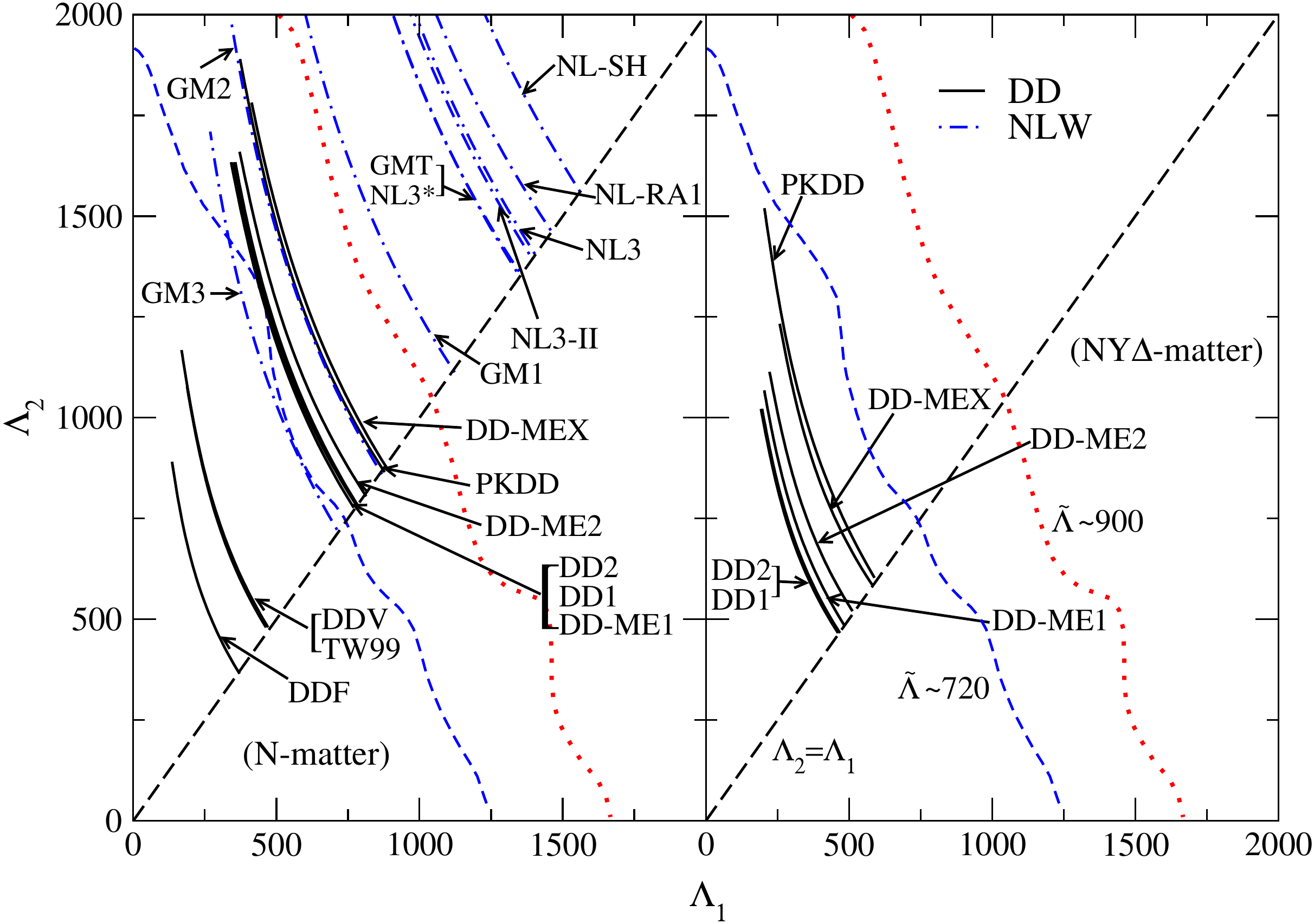}
\caption{Tidal deformability parameters $\Lambda_1$, $\Lambda_2$ corresponding to the binary components, $M_1,~M_2$ for GW170817 event with matter composition as, left panel: pure nucleonic matter, right panel: $\Delta$-baryon admixed hypernuclear and assuming a fixed chirp mass, $\mathcal{M}=1.188~M_{\odot}$. The solid curves denote the DD type parametrizations while the NLW type are denoted by dash-dotted curves. The dotted, short-dashed curves denote the $\tilde{\Lambda} \sim 900$ (TaylorF2), $720$ (PhenomPNRT) upper bounds at $90\%$ confidence level (low-spin priors) respectively \citep{LIGO_Virgo2017c,PhysRevX.9.011001}. The diagonal long-dashed line marks $\Lambda_1=\Lambda_2$ case.}
\label{fig-006}
\end{center}
\end{figure*}

We now look into the pressure bounds in lower and higher matter density regimes derived from GW170817 event data. Fig.-\ref{fig-003} shows that all the EOSs with pure nuclear matter satisfying mass-radius constraints also satisfy both the pressure bounds. It is to be noted that among NLW models, GM1, GM2 and GM3 satisfy only the lower bound at $n\sim 2n_0$ while they fail to satisfy the bound at $n\sim 6n_0$.
It has to be kept in mind that the estimated bound at $6n_0$ is more than the central pressures of compact stars involved in GW170817 event.

We have evaluated $\tilde{\Lambda}$ with the range of primary and secondary masses $1.36-1.60~M_{\odot}$ and $1.17-1.36~M_{\odot}$ respectively to provide the chirp mass, $\mathcal{M}=(M_1 M_2)^{3/5}(M_T)^{-1/5}=1.188~M_{\odot}$ where the total mass $M_T=M_1 + M_2$, is in the range $2.73-2.78~M_{\odot}$ for GW170817 event. 
In this work, we have considered the source properties to be circumscribing within $90\%$ credible intervals.
Fig.-\ref{fig-004} shows the $\tilde{\Lambda}$ variation with mass-ratio parameter ($q$). 
$\tilde{\Lambda}$ is found to be almost  independent of the mass asymmetry factor $q$ 
(refer to table-\ref{tab:2} for numerical results). In the left panel, curves are for pure nucleonic matter and in the right panel, they are for $\Delta$-admixed hypernuclear matter. For pure nucleonic matter, only GM2, GM3 models lie within the observational $\tilde{\Lambda}\sim 900$ bound among the NLW CDF models. However, they do not satisfy the lower bound constraint of maximum mass. 
In DD CDF models, all parametrizations considered in this work follow the upper bound of $900$. The stringent limit of $\tilde{\Lambda}\sim 720$ is observed to be satisfied by only TW99, DDV, DDF models which do not satisfy lower bound constraint for maximum mass. 
With $\Delta$-baryons coming into the picture, the models fulfil the latter stringent upper bound on $\tilde{\Lambda}$ as shown from the right panel of the figure.
Since the effect of hyperon inclusion on $\tilde{\Lambda}$ is similar to that of nucleons for the NSs with mass bounds obtained from GW170817 event, so they are not shown in fig.-\ref{fig-004}.
\begin{figure} 
  \begin{center}
\includegraphics[width=8.5cm,keepaspectratio ]{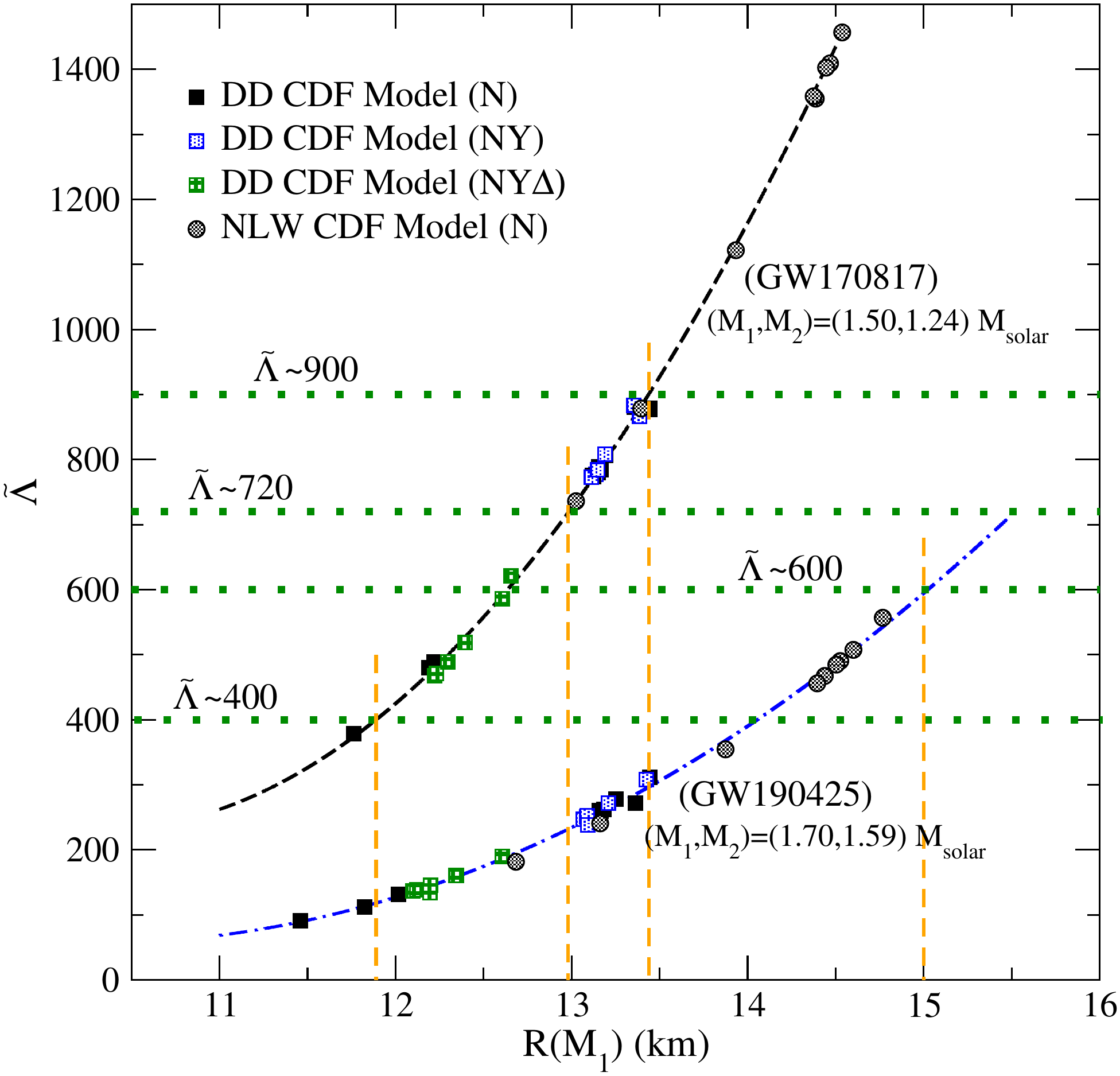}
\caption{Effective tidal deformability variation with radius of the binary's primary component for GW170817 and GW190425 events. The solid, dotted, stripped squares represent the DD parametrizations with pure nucleonic, baryon octet, $\Delta$-admixed hypernuclear matter respectively. The solid circles denote the NLW parametrizations with nucleonic matter case. The quadratic correlations for the GW170817, GW190425 cases are given by short-dashed and dash-dotted curves respectively. The horizontal dotted lines represent bounds on $\tilde{\Lambda}$ similar to figs.-\ref{fig-004},\ref{fig-005}. The vertical short-dashed lines mark the points where the quadratic fits intersect $\tilde{\Lambda}$ bounds. 
}
\label{fig-007}
\end{center}
\end{figure}

\begin{table*} 
\caption{Observational properties of various CDF models with different matter compositions. Here, $q$ represents the mass ratio of the secondary component ($M_2$) to primary one ($M_1$) involved in GW event. $C_{1.4}$ denotes the compactness parameter for a $1.4M_{\odot}$ NS.}
\centering
\begin{tabular}{cccccccccc}
\hline \hline
Matter & CDF Model & $M_{\text{max}}$ & $R_{1.4}$ & $\Lambda_{1.4}$ & $C_{1.4}$ & \multicolumn{2}{c}{$\tilde{\Lambda}~(q=0.8)$} & \multicolumn{2}{c}{$\tilde{\Lambda}~(q=1.0)$}  \\
composition & & ($M_{\odot}$) & (km) & & & GW170817 & GW190425 & GW170817 & GW190425 \\
\hline
& GM1 & 2.387 & 13.939 & 966.34 & 0.148 & 1109.39 & 355.77 & 1121.63 & 358.93 \\
& GM2 & 2.079 & 13.468 & 743.19 & 0.153 & 870.59 & 246.12 & 871.51 & 244.33 \\
& GM3 & 2.019 & 13.146 & 607.88 & 0.157 & 724.64 & 189.05 & 728.79 & 184.81 \\
& NL3 & 2.774 & 14.430 & 1222.72 & 0.143 & 1394.86 & 489.88 & 1410.87  & 495.58 \\
& NL3-II & 2.773 & 14.408 & 1217.59 & 0.143 & 1381.14 & 483.69 & 1401.64 & 494.33 \\
& NL-SH & 2.799 & 14.630 & 1368.94 & 0.141 & 1547.36 & 554.41 & 1563.07  & 565.93 \\
& NL-RA1 & 2.785 & 14.490 & 1278.05 & 0.143 & 1440.78 & 508.58 & 1465.28 & 518.21 \\
Pure & NL3* & 2.762 & 14.355 & 1181.41 & 0.144 & 1341.70 & 465.63 & 1358.48 & 476.51 \\
Nucleonic & GMT & 2.662 & 14.355 & 1177.91 & 0.144 & 1342.24 & 456.92 & 1359.93 & 462.36 \\
Matter & DD1 & 2.410 & 13.126 & 678.92 & 0.157 & 772.75 & 259.42 & 784.91 & 261.88 \\
& DD2 & 2.418 & 13.133 & 683.97 & 0.157 & 778.13 & 261.79 & 787.99 & 265.77 \\
& DD-ME1 & 2.443 & 13.086 & 672.37 & 0.158 & 765.14 & 261.09 & 776.04  & 266.31 \\
& DD-ME2 & 2.483 & 13.146 & 706.08 & 0.157 & 798.93 & 276.48 & 811.16  & 281.47 \\
& PKDD & 2.328 & 13.461 & 750.66 & 0.154 & 869.05 & 273.56 & 874.03 & 275.28 \\
& TW99 & 2.076 & 12.245 & 402.46 & 0.169 & 474.17 & 134.40 & 474.57 & 133.73 \\
& DDV & 1.929 & 12.360 & 398.67 & 0.167 & 485.52 & 117.40 & 479.76 & 112.95 \\
& DDF & 1.956 & 11.871 & 311.618 & 0.174 & 372.22 & 94.59 & 375.27 & 91.84 \\
& DD-MEX & 2.556 & 13.293 & 773.49 & 0.156 & 869.28 & 309.58 & 883.91  & 316.32 \\
\hline
& DD1 & 2.039 & 13.124 & 680.44 & 0.158 & 777.19 & 262.33 & 783.48 & 264.47 \\
& DD2 & 2.046 & 13.132 & 684.61 & 0.157 & 784.18 & 265.08 & 793.09 & 267.20 \\
Hypernuclear & DD-ME1 & 2.075 & 13.086 & 673.16 & 0.158 & 773.05 & 265.19 & 778.05 & 267.71 \\
Matter & DD-ME2 & 2.115 & 13.146 & 707.53 & 0.157 & 808.16 & 277.39 & 813.83 & 282.33 \\
& PKDD & 1.943 & 13.444 & 744.97 & 0.154 & 866.23 & 269.07 & 870.23 & 273.08 \\
& DD-MEX & 2.186 & 13.293 & 773.49 & 0.156 & 873.62 & 312.38 & 884.72 & 319.36 \\
\hline
& DD1 & 2.052 & 12.254 & 398.11 & 0.169 & 462.63 & 137.06 & 465.33 & 138.56 \\
& DD2 & 2.059 & 12.260 & 402.03 & 0.169 & 466.22 & 139.86 & 469.16 & 140.69 \\
$\Delta$-admixed & DD-ME1 & 2.085 & 12.320 & 418.19 & 0.168 & 485.03 & 146.52 & 486.65 & 148.15 \\
Hypernuclear & DD-ME2 & 2.126 & 12.400 & 444.20 & 0.167 & 514.21 & 160.55 & 517.89 & 163.58 \\
Matter & PKDD & 1.934 & 12.832 & 501.68 & 0.161 & 617.18 & 144.53 & 602.05 & 136.74 \\
& DD-MEX & 2.198 & 12.588 & 503.30 & 0.164 & 577.19 & 189.24 & 586.07 & 192.52 \\
\hline
\end{tabular}
\label{tab:2}
\end{table*} 

\begin{figure} 
  \begin{center}
\includegraphics[width=8.5cm,keepaspectratio ]{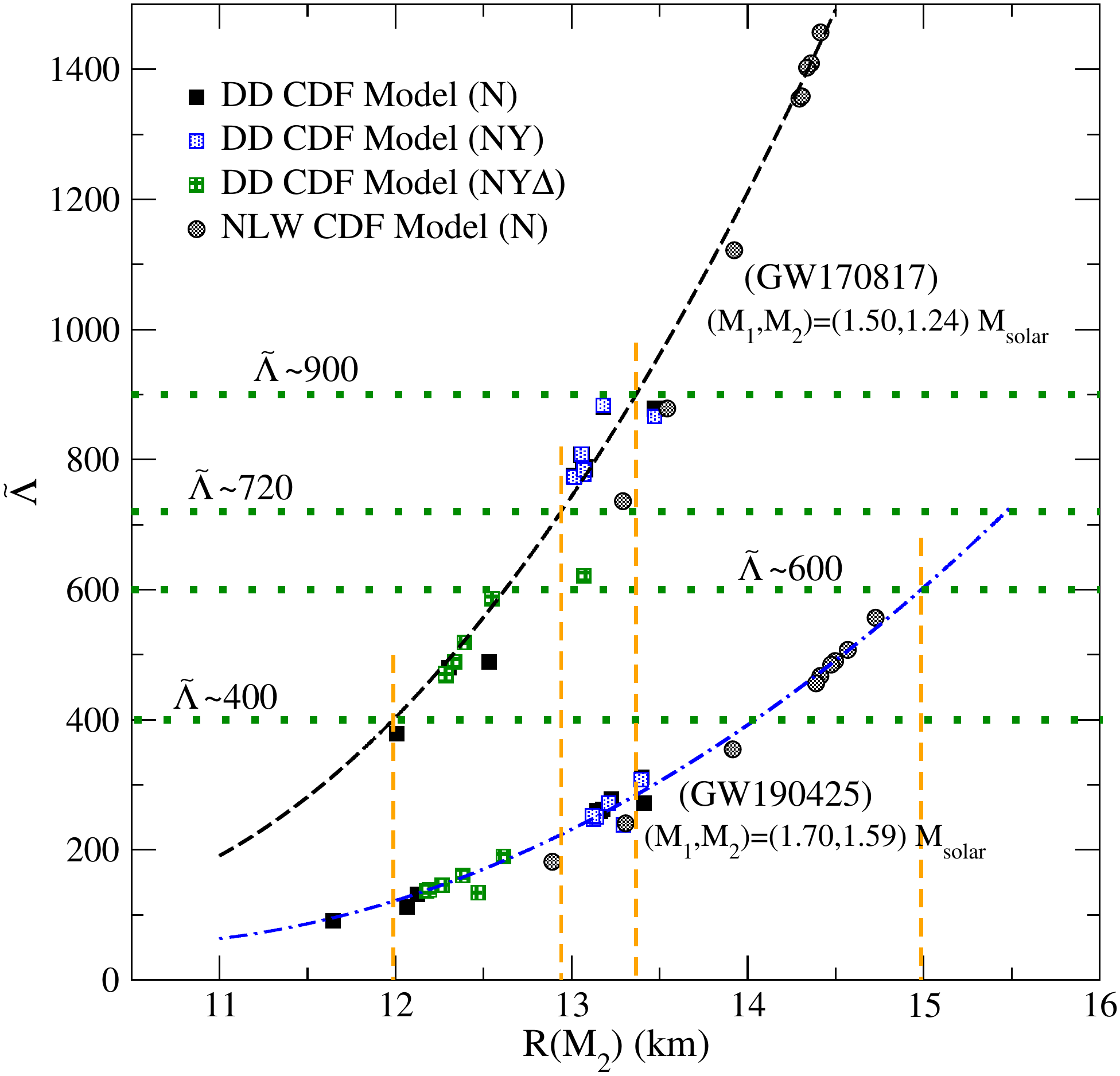}
\caption{Similar to fig.-\ref{fig-007} but for secondary components of GW170817 and GW190425 events. The solid, dotted, stripped squares represent the DD parametrizations with pure nucleonic, baryon octet, $\Delta$-admixed hypernuclear matter respectively. The solid circles denote the NLW parametrizations with nucleonic matter case. The quadratic correlations for the GW170817, GW190425 cases are given by short-dashed and dash-dotted curves respectively. $\tilde{\Lambda}$ constraints are similar to fig.-\ref{fig-007}.
}
\label{fig-011}
\end{center}
\end{figure}

Similar to fig.-\ref{fig-004}, the effective tidal deformability as a function of mass-asymmetry factor $q$ corresponding to the GW190425 event with a fixed chirp mass, $\mathcal{M}=1.43~M_{\odot}$ for all the parametrizations considered in this work is shown in fig.-\ref{fig-005} with only nucleonic matter. In this case, masses of the two compact stars are varied in the ranges $1.60 \leq M_1/M_{\odot} \leq 1.87$ (primary) and $1.46 \leq M_2/M_{\odot} \leq 1.69$ (secondary) \citep{2020ApJ...892L...3A}. Weak dependece of $\tilde{\Lambda}$ on $q$ can be inferred. It is observed that all the parametrizations satisfy the upper bound constraint on $\tilde{\Lambda}$ provided by GW190425 event data. Consequently this GW event does not provide enough information to put strict limits on constraining dense matter EOSs. Inclusion of $\Delta$-quartet will decrease $\tilde{\Lambda}$ compared to pure nucleonic case.

We next evaluate the tidal deformabilities ($\Lambda_1$, $\Lambda_2$) of binary components involved in GW170817 event with different matter compositions. For the evaluation of $\Lambda_1$ and $\Lambda_2$, we consider 
$\mathcal{M}=1.188~M_{\odot}$ where $M_T=2.73-2.78~M_{\odot}$. The masses of the two components are varied in $1.36 \leq M_1/M_{\odot} \leq 1.60$ (primary) and $1.17 \leq M_2/M_{\odot} \leq 1.36$ (secondary) ranges \citep{LIGO_Virgo2017c}. 
From the left panel of fig.-\ref{fig-006}, it is observed that for pure nucleonic matter the NLW CDF model parametrization GM1 do not lie within the $90\%$ probability contours of $\tilde{\Lambda}\sim 900$, although GM2 and GM3 do, while in case of DD CDF models, all the parametrizations lie inside these $\tilde{\Lambda}\sim 900$ contours. 
In the case of matter composition as $NY$, the tidal deformability is quite similar to the ones with pure nucleonic matter and hence not shown in fig.-\ref{fig-006}.
In case of $\Delta$-resonance admixed hypernuclear matter, for all the relevant EOSs $\Lambda_1$, $\Lambda_2$ falls well even within the  $\tilde{\Lambda}\sim 720$ probability contour (obtained from recent reanalysis) as shown in the right panel of the figure.
Table-\ref{tab:2} provides the numerical estimates of various observational properties of different CDF models considering matter composition to be purely nucleonic, hypernuclear and $\Delta$-resonance admixed hypernuclear matter.
\begin{figure} 
  \begin{center}
\includegraphics[width=8.5cm,keepaspectratio ]{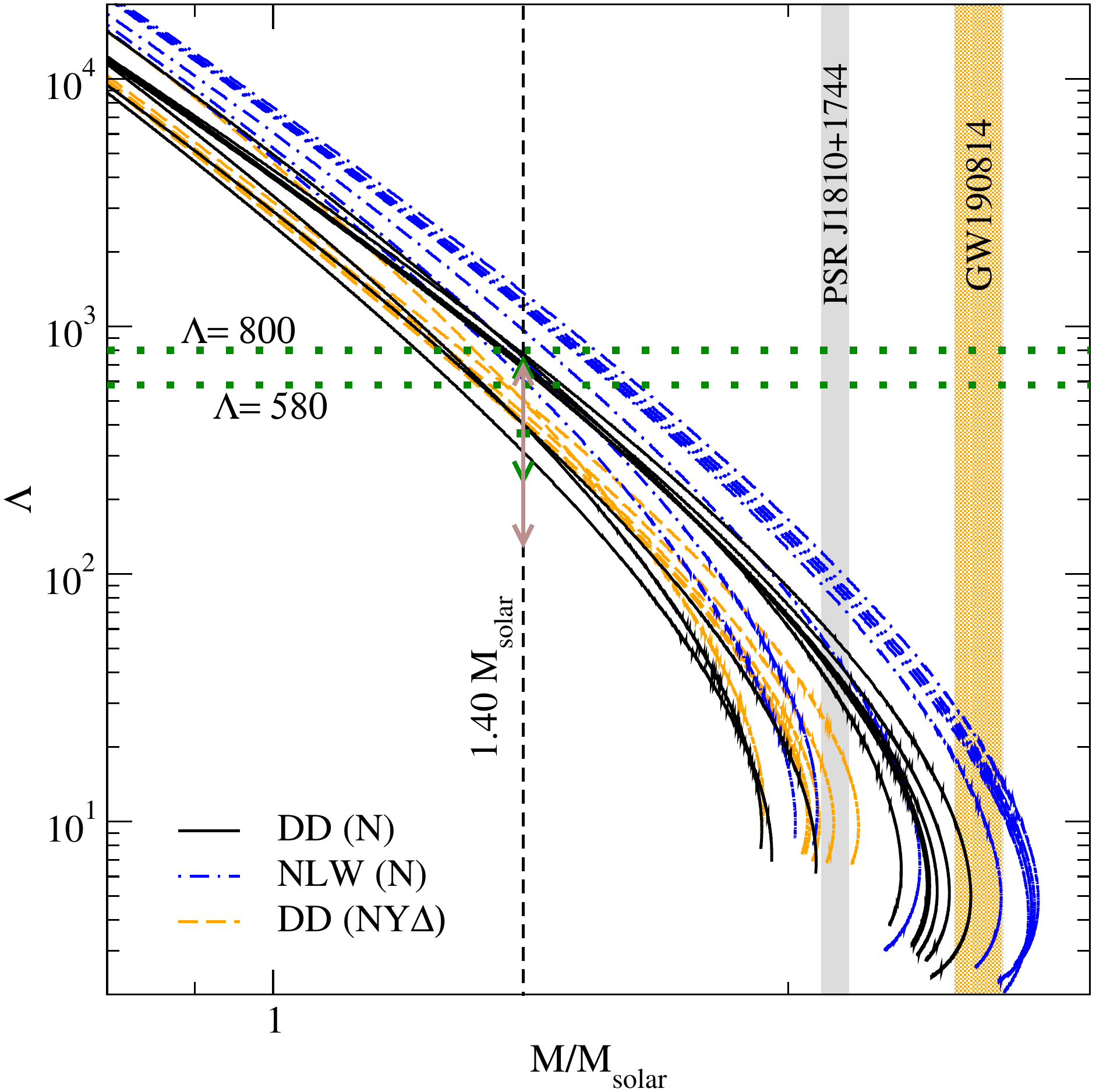}
\caption{Dimensionless tidal deformability ($\Lambda$) 
as a function of the NS mass ($M$) for CDF model 
parametrizations considered in this work 
(with $N$, $NY\Delta$-matter). DD, NLW type models with $N$-matter are represented similar to fig.-\ref{fig-001},\ref{fig-003}. While DD models with $NY\Delta$-matter are represented by long-dashed curves. The horizontal dotted lines denote the upper bounds on $\Lambda_{1.4}=800,580$ obtained from \citet{LIGO_Virgo2017c} and recent reanalysis \citet{LIGO_Virgo2018a} respectively (GW170817 event). The vertical range denotes joint constraints from NICER observations (PSR J$0030+0451$) $\&$ GW170817 event \citep{Jiang_2020} and another obtained implementing Bayesian analysis \citep{2021EPJA...57...31L} for a $1.4~M_{\odot}$ NS. The various astrophysical constraints are similar as in fig.-\ref{fig-001}.}
\label{fig-008}
\end{center}
\end{figure}

Next we attempt to restrict radius of compact stars by evaluating $\tilde{\Lambda}$ with particular star combinations for both the events considering different parametrization models. For the event GW170817, we have taken $M_1=1.50~M_{\odot}$ and $M_2=1.24~M_{\odot}$ and for the event GW190425 we have taken $M_1=1.70~M_{\odot}$ and $M_2=1.59~M_{\odot}$ where, $M_1$, $M_2$ correspond to primary and secondary components respectively. We plot the values of $\tilde{\Lambda}$ with radius of the primary star for respective EOSs in fig.-\ref{fig-007}. Tight correlations between $\tilde{\Lambda}$ and $R(M_1)$ are given by the fits
\begin{equation*}
\begin{aligned}
\tilde{\Lambda}^{\text{(GW170817)}}_{\text{fit}} & = 7571 - 1423R(M_1) + 68.93(R(M_1))^2, \\
\tilde{\Lambda}^{\text{(GW190425)}}_{\text{fit}} & = 2638 - 501.2R(M_1) + 24.34(R(M_1))^2 ,
\end{aligned}
\end{equation*}
with maximum deviations, $(|\tilde{\Lambda}_{\text{fit}}-\tilde{\Lambda}|/\tilde{\Lambda})\sim 2.65\%,~7.45\%$, 
$\chi^2 = \sum_{i}^{\mathcal{N}}\left[(\tilde{\Lambda}_{\text{fit}}^i-\tilde{\Lambda}^i)^2/\tilde{\Lambda}^i \right] = 4.16,~7.16$ with $\mathcal{N}=30$ EOS models and coefficients of determination, $\mathcal{R}^2 = 1 - SS_{\text{res}}/SS_{\text{total}}$
$\sim 0.999,~0.996$ for GW170817, GW190425 event respectively. Here, $SS_{\text{res}}=\sum_i (\tilde{\Lambda}_i-\tilde{\Lambda}_i^{\text{fit}})^2$, $SS_{\text{total}}=\sum_i (\tilde{\Lambda}_i-\bar{\tilde{\Lambda}}_i)^2$ are sum of squares of the residual errors and squared error of the mean line respectively.
The point where the curves cross $\tilde{\Lambda}$ bounds corresponds to limits on primary component's radius and hence the EOSs. 
The figure shows that the upper bound on $\tilde{\Lambda} \sim 900$ results in radius $\leq 13.44$ km, which excludes certain NLW models except for GM2 and GM3 EOSs. On the other hand, if we consider the upper bound on $\tilde{\Lambda} \sim 720$, the radius bound $\leq 12.98$ km not only excludes all the NLW EOSs but also matter without $\Delta$ resonances with DD parametrizations. From the lower bound of $\tilde{\Lambda} \sim 400$, the lower bound on radius $\geq 11.89$ km which excludes dense matter composed of only nucleons with DDF (DD type) parametrization. 
From the observation of GW190425, the upper bound on $\tilde{\Lambda}\sim 600$ provides the upper bound on the primary component's radius $\leq 15.00$ km, which gives no limit on the EOSs.
Similar to fig.-\ref{fig-007}, effective tidal deformability as a function of secondary component's radius in GW170817 and GW190425 events is shown in fig.-\ref{fig-011}. The correlation fits between $\tilde{\Lambda}$ and $R(M_2)$ are given by
\begin{equation*}
\begin{aligned}
\tilde{\Lambda}^{\text{(GW170817)}}_{\text{fit}} & = 6249 - 1251R(M_2) + 63.62(R(M_2))^2, \\
\tilde{\Lambda}^{\text{(GW190425)}}_{\text{fit}} & = 2783 - 527.5R(M_2) + 25.48(R(M_2))^2 ,
\end{aligned}
\end{equation*}
with maximum deviations of $\sim 17.76\%,~19.49\%$, $\chi^2 = 54.49,~17.44$ and $\mathcal{R}^2 \sim 0.983,~0.988$ for GW170817 and GW190425 events respectively. The upper bounds on radius for the secondary components in GW170817 case are estimated to be $13.37$ km ($\tilde{\Lambda} \sim 900$) and $12.94$ km ($\tilde{\Lambda} \sim 720$) while the lower bound is evaluated to be $11.99$ km ($\tilde{\Lambda} \sim 400$). In case of GW190425 case, radius of the secondary component, $R_{1.59} \leq 14.99$ km ($\tilde{\Lambda} \sim 600$).
From both the correlations it can be inferred that the radius bounds on NSs involved in GW events are approximately $12 \leq R_{*}/\text{km} \leq 13$ and $R_{*} \leq 15$ km in GW170817 and GW190425 events respectively.

\begin{table} 
\caption{Threshold densities denoted by $n_{u}^{Y}$ (in units of $n_0$) for onset of hyperons in hypernuclear dense matter for various DD CDF models. $n_{1.4}^{c(Y)}$ represents the central number density for a $1.4M_{\odot}$ NS with hypernuclear matter composition. $\Lambda_{1.4}^{N}$, $\Lambda_{1.4}^{NY}$ are the dimensionless tidal deformability of a $1.4M_{\odot}$ NS with nucleonic and hypernuclear matter respectively.}
\centering
\begin{tabular}{cccc}
\hline \hline
CDF Model & $n_{1.4}^{c(Y)}~(n_0)$ & $n_{u}^{Y}~(n_0)$ & $\Lambda_{1.4}^{N}/\Lambda_{1.4}^{NY}$ \\
\hline
DD1 & 2.42 & 2.26 & 0.998 \\
DD2 & 2.40 & 2.25 & 0.999 \\
DD-ME1 & 2.33 & 2.23 & 0.999 \\
DD-ME2 & 2.25 & 2.21 & 0.998 \\
PKDD & 2.48 & 2.15 & 1.01 \\
DD-MEX & 2.13 & 2.15 & 1.00 \\
\hline
\end{tabular}
\label{tab:3}
\end{table} 

Fig.-\ref{fig-008} depicts the dimensionless tidal deformability parameter as a function of NS mass evaluated from the DD and NLW type models with matter composition to be nucleonic and $\Delta$-commixed baryon octet. It is observed that among the NLW CDF models, GM2 and GM3 parametrizations fulfil the $\Lambda_{1.4}=800$ upper bound (with $N$-matter), while they fail to satisfy the recent $\Lambda_{1.4}=580$ bound. Other NLW parametrizations produce larger radii NSs and, as a result, are more inclined to be easily deformable since $\lambda\sim R^5$ (i.e. higher tidal deformability values). Hence no NLW model considered in this work satisfy the mass and tidal deformability constraints simultaneously. In the case of DD CDF models (with $N$-matter), all the models satisfy the upper bound constraint on $\Lambda_{1.4}=800$. However, except TW99, DDV and DDF parametrizations none of them fulfil the upper bound ($\Lambda_{1.4}=580$) obtained from reanalysis of GW170817 event data. 
No coupling parameter set considered in this work is seen to satisfy the more strict constraints of $\Lambda_{1.4}=580$ and maximum mass simultaneously with pure nucleonic matter.
Another joint constraint from NICER (PSR J$0030+0451$) and GW170817 data sets an upper bound on $\Lambda_{1.4}=730$. 
Recent constraint on $\Lambda_{1.4}$ obtained from Bayesian analysis provides an upper bound of $686$.
DD CDF models (DD1, DD2, DD-ME1, DD-ME2) are observed to satisfy these criteria inclusive with the lower bound on $M_{\text{max}}$
(see table-\ref{tab:2} for numerical results). It should be noted, as evident from table-\ref{tab:3} appearance of heavier baryons are inevitable in a $1.4$ $M_\odot$ star with all DD CDF parametrizations except DD-MEX.

\begin{figure} 
  \begin{center}
\includegraphics[width=8.5cm,keepaspectratio ]{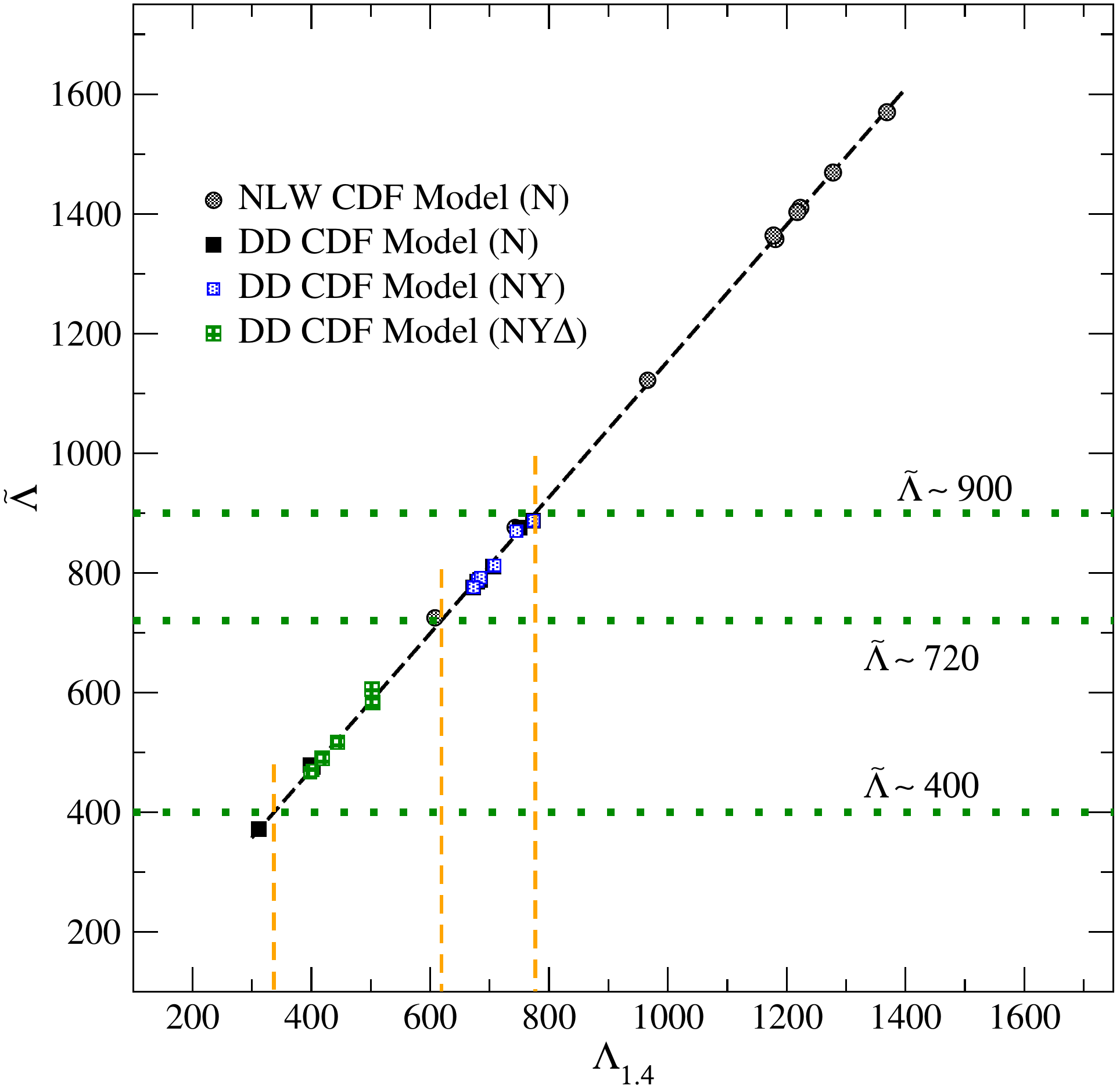}
\caption{Correlation between weighted average tidal deformability and tidal deformability for $1.4~M_{\odot}$ NSs. The short-dashed line denotes the best fit for the EOS data sets. The bounds on ${\tilde{\Lambda}}\sim 900,720$ (upper), $400$ (lower) corresponding to GW170817 event are represented by the horizontal dotted lines similar to fig.-\ref{fig-007}. The vertical short-dashed lines mark the points where the linear fit intersects $\tilde{\Lambda}$ bounds. Parametrization models involved are similar as in fig.-\ref{fig-007}.}
\label{fig-013}
\end{center}
\end{figure}
Fig.-\ref{fig-013} displays the tight correlation between weighted average $\tilde{\Lambda}$ and $\Lambda_{1.4}$ tidal deformability for GW170817 event data. In this case, we have considered $M_1=1.40~M_{\odot}$, $M_2=1.33~M_{\odot}$ with $\mathcal{M}=1.1878~M_{\odot}$. The tight linear correlation between $\tilde{\Lambda}$ and $\Lambda_{1.4}$ is given by
\begin{equation*}
\begin{aligned}
\tilde{\Lambda}^{\text{(GW170817)}}_{\text{fit}} & = 16.28 + 1.138\Lambda_{1.4},
\end{aligned}
\end{equation*}
with maximum deviation of $\sim 3.11\%$, $\chi^2=2.39$ and $\mathcal{R}^2 \sim 0.999$. The upper bounds on $\Lambda_{1.4}$ are deduced to be $777,~619$ corresponding to $\tilde{\Lambda}\sim 900,~720$ respectively. While the lower bound on $\Lambda_{1.4}$ based on $\tilde{\Lambda}\sim 400$ (AT2017gfo) is estimated to be $337$. The upper bound on $\tilde{\Lambda}$ favours the DD parametrizations as evident from fig.-\ref{fig-013}.

\begin{figure} 
  \begin{center}
\includegraphics[width=8.50cm,keepaspectratio ]{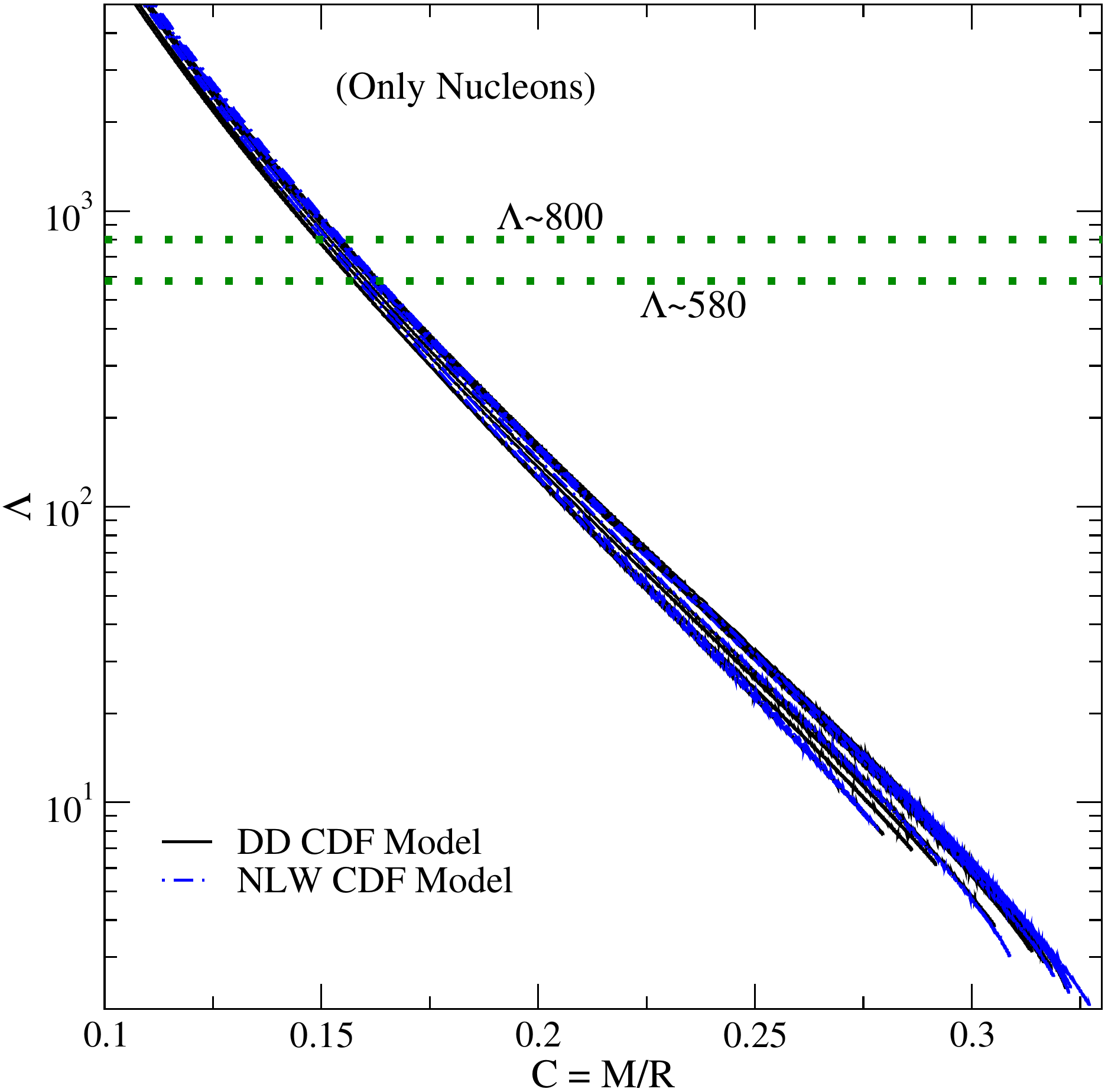}
\caption{Dimensionless tidal deformability as a function of compactness parameter ($C$) for an isolated NS (with $N$-matter) for the CDF parametrizations implemented in this work. The bounds on ${\Lambda}\sim 800,580$ corresponding to GW170817 event are represented by the horizontal dotted lines similar to fig.-\ref{fig-008}.}
\label{fig-009}
\end{center}
\end{figure}

\begin{figure} 
  \begin{center}
\includegraphics[width=8.5cm,keepaspectratio ]{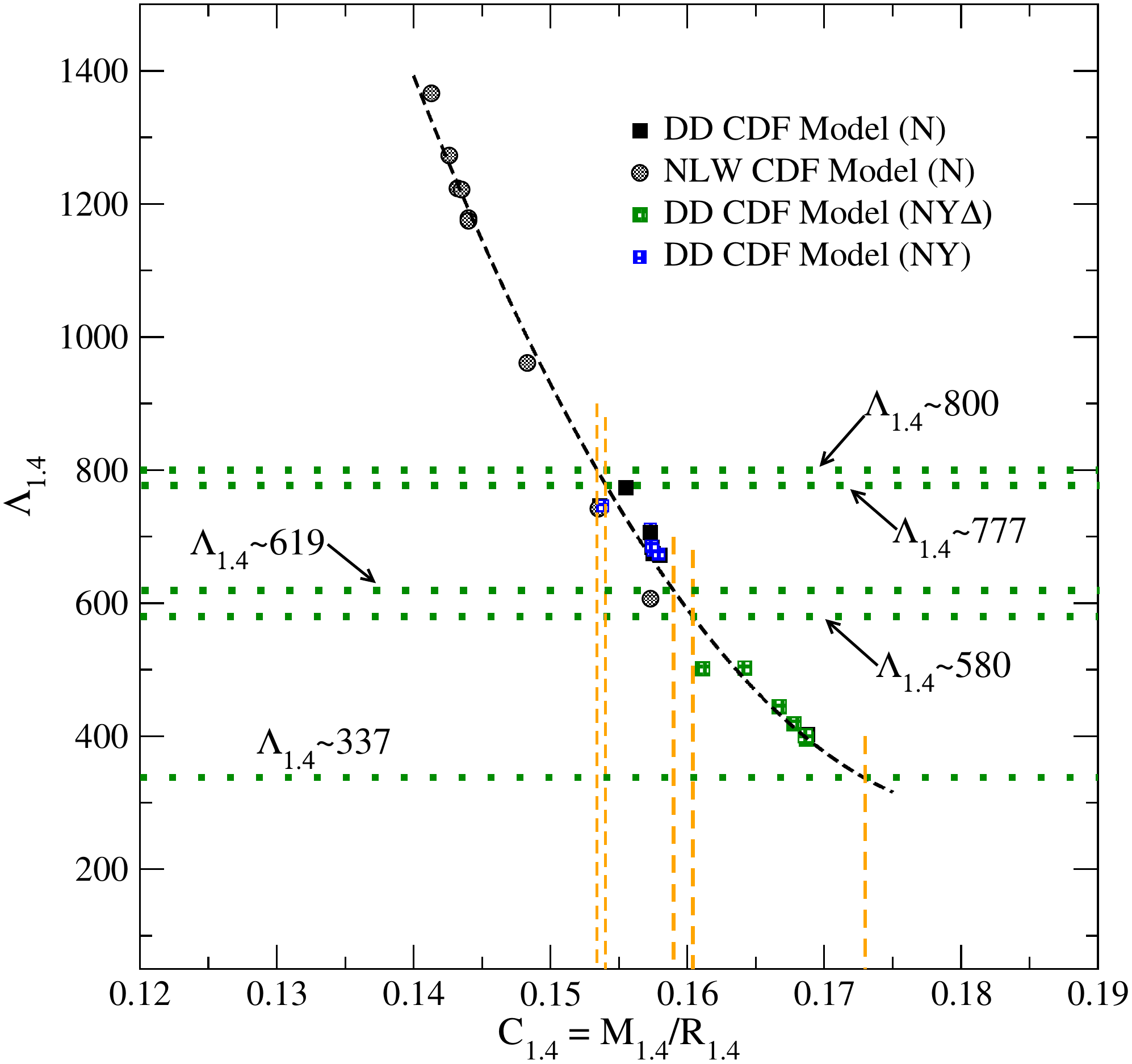}
\caption{Correlation between tidal deformability and compactness parameter for $1.4~M_{\odot}$ NSs. The short-dashed line denotes the best fit for the EOS data sets. The bounds on ${\Lambda_{1.4}}$ corresponding to GW170817 event are represented by the horizontal dotted lines similar to figs.-\ref{fig-008},\ref{fig-009} in addition to ones deduced from fig.-\ref{fig-013}. The vertical short-dashed lines mark the points where the quadratic fit intersects $\Lambda_{1.4}$ bounds. Parametrization models involved are similar as in fig.-\ref{fig-007}.}
\label{fig-010}
\end{center}
\end{figure}

We have shown the dependence of $\Lambda$ on compactness of the star for different parametrizations (with $N$-matter) in fig.-\ref{fig-009}. From the figure, it is clear that the dependence of $\Lambda$ on compactness is almost independent of EOSs. 
This relates with the result from \citet{2013PhRvD..88b3007M}.
Then in fig.-\ref{fig-010}, we plot the $\Lambda$ for an isolated NS of mass $1.4 M_\odot$ which shows a general trend with almost all EOSs. In order to find lower bound on the compactness parameter of a $1.4~M_{\odot}$ NS ($C_{1.4}$), we compute the correlations between $\Lambda_{1.4}$ and $C_{1.4}$ 
as $\Lambda_{1.4}/10^4 = 2.094 - 22.67 C_{1.4} +62.2 C^{2}_{1.4}$. 
The maximum deviation estimated is $\sim 11.94\%$ with $\chi^2=41.03$, $\mathcal{R}^2$ corresponding to $0.989$. The stiffer NLW type parametrization models produce compact stars with larger mass and radius leading to higher tidal deformabilities.
Obviously, these stiffer EOSs do not fulfil the upper limit of $\Lambda_{1.4} \leq 800$. Softer parametrizations GM2 and GM3 satisfy the upper limit of $\Lambda$.
On the other hand, DD type parametrizations considered in this work generate compact stars with $\Lambda_{1.4} \leq 800$. Inclusion of $\Delta$-resonances reduces $R_{1.4}$ resulting in increase of $C_{1.4}$ keeping $\Lambda$ less than its upper limit. 
From fig.-\ref{fig-010}, it can be inferred that for a $1.4~M_\odot$ NS, the lower bound in compactness is $0.153(0.154)$  
corresponding to $\Lambda_{1.4}\sim 800(777)$ and it is $0.160(0.159)$ following the stringent upper bound of $\Lambda_{1.4}\sim 580(619)$. 
It also shows that the upper bound in compactness for a $1.4~M_\odot$ NS is $0.173$ following the lower bound $\Lambda_{1.4}\sim 337$. 
The points which lie away from the correlation fit are from GM2, GM3, PKDD coupling models. 
Based on the derived bounds of $C_{1.4}$, the range of $R_{1.4}$ is found to be $11.95-13.00$ km ($C_{1.4}\sim 0.159-0.173$) and $11.95-13.42$ km ($0.154-0.173$).
The estimated bounds on $R_{1.4}$ satisfy the range $11.5 \leq R_{1.4}/\text{km} \leq 13.6$ as reported in \citet{2006PhLB..642..436L} with the latter constrained from terrestrial experimental data.

In addition to heavier baryons, meson such as (anti)kaon condensations may also come into the picture with rising energy density towards the interior of NSs. \citet{PhysRevD.102.123007, PhysRevD.103.063004} reported that these meson condensations affect the lowering of maximum mass NS configurations and do not have any significant reduction in radii of $\sim 1.4~M_{\odot}$ NSs compared to pure hadronic matter cases.

\section{Conclusions and Outlook}\label{sec:summary}

The BNS mergers provide intriguing information to constrain various theoretical formulations of dense nuclear matter EOSs. In this work, we have analysed two schemes of coupling parametrizations and tried to constrain purely nucleonic, hypernuclear, and $\Delta$-baryon admixed hypernuclear matter EOSs within the CDF framework model via employing GW and other astrophysical observations.

Confronting the coupling parametrizations considered in this work with the recent bounds on nuclear saturation properties infers that the NLW parameter sets fail to satisfy these empirical ranges while DD coupling models (except PKDD) are seen to fulfil these bounds. In addition, imposing the lower limit of maximum mass constraint rules out the relatively softer EOSs viz. GM2, GM3 among NLW models and DDF, DDV, TW99 among DD CDF models. On the other hand, very stiff NLW EOSs, except GM1 fail to satisfy the measured radius range from observation of PSR J$0030+0451$. In the core of NSs, with enhanced matter density, the appearance of heavier baryons are inevitable. Their appearance softens the matter, excluding some more parametrization viz. DD1, DD2, PKDD among DD CDF models and only admissible NLW model parametrization GM1. Even with pure nucleonic matter, the parametrization GM1 is ruled out because it is not as soft as to provide $\tilde{\Lambda}\le 900$ from GW170817 event. Hence, none of the parameter sets within NLW model considered in this work can satisfy all the observational constraints simultaneously. The upper bound on $\tilde{\Lambda}\le 900$ allow all the parametrizations of DD CDF models, although the strict upper limit $\tilde{\Lambda}\le 720$ does not allow any parametrization with pure nuclear matter maintaining lower bound constraints of maximum mass. However, the appearance of heavier non-strange baryons allows all the DD CDF parametrizations to lie within the stringent upper limit.

The advent of heavier baryons (hyperons) leads to softening of EOSs, which in turn reduce the NS maximum masses values by $\sim 0.38~M_{\odot}$ than that of pure nucleonic cases. 
The effect on dimensionless tidal deformability ($\Lambda_{1.4}$) due to the incorporation of hyperons is very marginal.
It can be seen that the central number densities of $1.4M_{\odot}$ NSs are quite close to the onset of hyperons. This attributes to the marginal changes in $\Lambda_{1.4}$. 
However, the appearance of $\Delta$-resonances in dense matter is worthy of mentioning. 
From fig.-\ref{fig-008}, it can be seen that the value of $\Lambda_{1.4}$ lies above the stringent $580$ upper bound for the stiffer DD CDF EOSs for pure nucleonic matter. With the advent of $\Delta$-quartet, $\Lambda_{1.4}$ reduces sufficiently and moves below $580$ for all stiff DD CDF parametrizations considered in this work. This relates with the results from \citet{Li2019ApJ} and \citet{2021PhLB..81436070R}. 
$\Delta$-baryon admixed hypernuclear NSs are observed to have smaller radius values in comparison to the NSs with only baryon octet particle spectrum (refer to table-\ref{tab:2}). This is due to the early onset of $\Delta^-$ baryon, which relatively softens the EOSs at lower densities.

\cite{2021ApJ...915L...5A} recently reported the observation of GWs from two NS-BH coalescences (GW200105 and GW200115). The masses of the NSs involved are deduced to be $1.9^{+0.3}_{-0.2}~M_{\odot}$, $1.5^{+0.7}_{-0.3}~M_{\odot}$ for the GW200105 and GW200115 events respectively. However, no constraints on tidal deformation of the secondary components have been reported. The coupling parametrizations considered in this work satisfy the NS mass constraints set by these recent GW events. 

Based on the coupling data sets contemplated in this work and observed strict restrictions on $\tilde{\Lambda}$, a radius range of the NSs involved in GW170817 event is deduced to be around $12 \leq R_{*}/\text{km} \leq 13$ and that with GW190425 event's data is found to be $R_* \leq 15.00$ km. The lower bound on $\tilde{\Lambda}$ considered in this work is evaluated from the electromagnetic counterpart (kilonova) signal in GW170817 event.
The strong correlation between weighted average $\tilde{\Lambda}$ and $\Lambda_{1.4}$ sets bounds as $337 \leq \Lambda_{1.4} \leq 619$ corresponding to $400 \leq \tilde{\Lambda} \leq 720$.
Furthermore, similar analysis with strict bounds on $\Lambda_{1.4}$ translate to compactness parameter value of a $1.4~M_{\odot}$ NS to be in the range $0.159 \leq C_{1.4} \leq 0.173$. This yields $R_{1.4}$ in the range $11.95-13.00$ km.

Another inference from the GW observational constraints is the possible existence of quark matter in the interior of NSs. Many works \citep{Nandi_2018, 2019ApJ...877..139G,2018PhRvD..97h4038P, 2019MNRAS.489.4261M, 2021EPJST.tmp....5N, PhysRevC.103.055814} reported hybrid star configurations consistent with the strict upper bounds on $\tilde{\Lambda}$, $\Lambda_{1.4}$ and massive NS observations. Since the knowledge regarding high-density quark matter is still inadequate, we could explore and constrain the parameter space of the functionals involved in describing the quark matter behaviour at such density regimes based on GW events and recent NICER data. Further analysis on the aspect is beyond the scope of this work and will be discussed in future studies.

\section*{Acknowledgements}
The authors thank the anonymous referee for constructive comments that significantly contributed to enhancing the manuscript's quality.
The authors acknowledge financial support from Science and Engineering Research Board, Department of Science and Technology, Government of India through Project No. EMR/2016/006577 and Ministry of Education, Government of India. VBT and MS are also thankful to Sarmistha Banik for fruitful discussions.

\section*{Data Availability}
Data sharing not applicable to this article as no data sets were generated during this study.


\bibliographystyle{mnras}
\bibliography{references} 

\begin{thebibliography}{}
\makeatletter
\relax
\def\mn@urlcharsother{\let\do\@makeother \do\$\do\&\do\#\do\^\do\_\do\%\do\~}
\def\mn@doi{\begingroup\mn@urlcharsother \@ifnextchar [ {\mn@doi@}
  {\mn@doi@[]}}
\def\mn@doi@[#1]#2{\def\@tempa{#1}\ifx\@tempa\@empty \href
  {http://dx.doi.org/#2} {doi:#2}\else \href {http://dx.doi.org/#2} {#1}\fi
  \endgroup}
\def\mn@eprint#1#2{\mn@eprint@#1:#2::\@nil}
\def\mn@eprint@arXiv#1{\href {http://arxiv.org/abs/#1} {{\tt arXiv:#1}}}
\def\mn@eprint@dblp#1{\href {http://dblp.uni-trier.de/rec/bibtex/#1.xml}
  {dblp:#1}}
\def\mn@eprint@#1:#2:#3:#4\@nil{\def\@tempa {#1}\def\@tempb {#2}\def\@tempc
  {#3}\ifx \@tempc \@empty \let \@tempc \@tempb \let \@tempb \@tempa \fi \ifx
  \@tempb \@empty \def\@tempb {arXiv}\fi \@ifundefined
  {mn@eprint@\@tempb}{\@tempb:\@tempc}{\expandafter \expandafter \csname
  mn@eprint@\@tempb\endcsname \expandafter{\@tempc}}}

\bibitem[\protect\citeauthoryear{Abbott, Abbott, Abbott, Acernese, Ackley
  et~al.}{Abbott et~al.}{2017a}]{LIGO_Virgo2017c}
Abbott B.~P.,  Abbott R.,  Abbott T.~D.,  Acernese F.,  Ackley K.,   et~al.,
  2017a, PhRvL, 119, 161101

\bibitem[\protect\citeauthoryear{Abbott, Abbott, Adhikari, Ananyeva, Anderson
  et~al.}{Abbott et~al.}{2017b}]{LIGO_Virgo2017a}
Abbott B.~P.,  Abbott R.,  Adhikari R.~X.,  Ananyeva A.,  Anderson S.~B.,
  et~al., 2017b, ApJL, 848, L12

\bibitem[\protect\citeauthoryear{Abbott, Abbott, Abbott, Acernese, Ackley
  et~al.}{Abbott et~al.}{2017c}]{LIGO_Virgo2017b}
Abbott B.~P.,  Abbott R.,  Abbott T.~D.,  Acernese F.,  Ackley K.,   et~al.,
  2017c, ApJL, 848, L13

\bibitem[\protect\citeauthoryear{Abbott, Abbott, Abbott, Acernese, Ackley
  et~al.}{Abbott et~al.}{2018}]{LIGO_Virgo2018a}
Abbott B.~P.,  Abbott R.,  Abbott T.~D.,  Acernese F.,  Ackley K.,   et~al.,
  2018, PhRvL, 121, 161101

\bibitem[\protect\citeauthoryear{Abbott, Abbott, Abbott  et~al.}{Abbott
  et~al.}{2019}]{PhysRevX.9.011001}
Abbott B.~P.,  Abbott R.,  Abbott T.~D.,   et~al., 2019, \mn@doi [Phys. Rev. X]
  {10.1103/PhysRevX.9.011001}, 9, 011001

\bibitem[\protect\citeauthoryear{{Abbott} et~al.,}{{Abbott}
  et~al.}{2020a}]{2020ApJ...892L...3A}
{Abbott} B.~P.,  et~al., 2020a, \mn@doi [\apjl] {10.3847/2041-8213/ab75f5},
  \href {https://ui.adsabs.harvard.edu/abs/2020ApJ...892L...3A} {892, L3}

\bibitem[\protect\citeauthoryear{{Abbott} et~al.,}{{Abbott}
  et~al.}{2020b}]{2020ApJ...896L..44A}
{Abbott} R.,  et~al., 2020b, \mn@doi [\apjl] {10.3847/2041-8213/ab960f}, \href
  {https://ui.adsabs.harvard.edu/abs/2020ApJ...896L..44A} {896, L44}

\bibitem[\protect\citeauthoryear{{Abbott} et~al.,}{{Abbott}
  et~al.}{2021}]{2021ApJ...915L...5A}
{Abbott} R.,  et~al., 2021, \mn@doi [\apjl] {10.3847/2041-8213/ac082e}, \href
  {https://ui.adsabs.harvard.edu/abs/2021ApJ...915L...5A} {915, L5}

\bibitem[\protect\citeauthoryear{{Akmal}, {Pandharipande}  \&
  {Ravenhall}}{{Akmal} et~al.}{1998}]{1998PhRvC..58.1804A}
{Akmal} A.,  {Pandharipande} V.~R.,   {Ravenhall} D.~G.,  1998, \mn@doi [\prc]
  {10.1103/PhysRevC.58.1804}, \href
  {https://ui.adsabs.harvard.edu/abs/1998PhRvC..58.1804A} {58, 1804}

\bibitem[\protect\citeauthoryear{{Antoniadis} et~al.,}{{Antoniadis}
  et~al.}{2013}]{2013Sci...340..448A}
{Antoniadis} J.,  et~al., 2013, \mn@doi [Science] {10.1126/science.1233232},
  \href {https://ui.adsabs.harvard.edu/abs/2013Sci...340..448A} {340, 448}

\bibitem[\protect\citeauthoryear{Arzoumanian et~al.}{Arzoumanian
  et~al.}{2018}]{Arzoumanian_2018}
Arzoumanian Z.,  et~al., 2018, \mn@doi [ApJ Suppl.] {10.3847/1538-4365/aab5b0},
  235, 37

\bibitem[\protect\citeauthoryear{{Baillot d'Etivaux}, {Guillot}, {Margueron},
  {Webb}, {Catelan}  \& {Reisenegger}}{{Baillot d'Etivaux}
  et~al.}{2019}]{2019ApJ...887...48B}
{Baillot d'Etivaux} N.,  {Guillot} S.,  {Margueron} J.,  {Webb} N.,  {Catelan}
  M.,   {Reisenegger} A.,  2019, \mn@doi [\apj] {10.3847/1538-4357/ab4f6c},
  \href {https://ui.adsabs.harvard.edu/abs/2019ApJ...887...48B} {887, 48}

\bibitem[\protect\citeauthoryear{{Banik} \& {Bandyopadhyay}}{{Banik} \&
  {Bandyopadhyay}}{2001}]{2001PhRvC..63c5802B}
{Banik} S.,  {Bandyopadhyay} D.,  2001, \mn@doi [\prc]
  {10.1103/PhysRevC.63.035802}, \href
  {https://ui.adsabs.harvard.edu/abs/2001PhRvC..63c5802B} {63, 035802}

\bibitem[\protect\citeauthoryear{{Bao} \& {Shen}}{{Bao} \&
  {Shen}}{2014}]{2014PhRvC..89d5807B}
{Bao} S.~S.,  {Shen} H.,  2014, \mn@doi [\prc] {10.1103/PhysRevC.89.045807},
  \href {https://ui.adsabs.harvard.edu/abs/2014PhRvC..89d5807B} {89, 045807}

\bibitem[\protect\citeauthoryear{{Baym}, {Pethick}  \& {Sutherland}}{{Baym}
  et~al.}{1971}]{1971ApJ...170..299B}
{Baym} G.,  {Pethick} C.,   {Sutherland} P.,  1971, \mn@doi [\apj]
  {10.1086/151216}, \href
  {https://ui.adsabs.harvard.edu/abs/1971ApJ...170..299B} {170, 299}

\bibitem[\protect\citeauthoryear{Baym, Hatsuda, Kojo, Powell, Song  \&
  Takatsuka}{Baym et~al.}{2018}]{Baym_2018}
Baym G.,  Hatsuda T.,  Kojo T.,  Powell P.~D.,  Song Y.,   Takatsuka T.,  2018,
  \mn@doi [Rep. Prog. Phys.] {10.1088/1361-6633/aaae14}, 81, 056902

\bibitem[\protect\citeauthoryear{Binnington \& Poisson}{Binnington \&
  Poisson}{2009}]{PhysRevD.80.084018}
Binnington T.,  Poisson E.,  2009, \mn@doi [Phys. Rev. D]
  {10.1103/PhysRevD.80.084018}, 80, 084018

\bibitem[\protect\citeauthoryear{{Biswas}}{{Biswas}}{2021}]{2021arXiv210502886B}
{Biswas} B.,  2021, arXiv e-prints, \href
  {https://ui.adsabs.harvard.edu/abs/2021arXiv210502886B} {p. arXiv:2105.02886}

\bibitem[\protect\citeauthoryear{{Boguta} \& {Bodmer}}{{Boguta} \&
  {Bodmer}}{1977}]{1977NuPhA.292..413B}
{Boguta} J.,  {Bodmer} A.~R.,  1977, \mn@doi [\nphysa]
  {10.1016/0375-9474(77)90626-1}, \href
  {https://ui.adsabs.harvard.edu/abs/1977NuPhA.292..413B} {292, 413}

\bibitem[\protect\citeauthoryear{{Bombaci}, {Drago}, {Logoteta}, {Pagliara}  \&
  {Vidana}}{{Bombaci} et~al.}{2020}]{2020arXiv201001509B}
{Bombaci} I.,  {Drago} A.,  {Logoteta} D.,  {Pagliara} G.,   {Vidana} I.,
  2020, arXiv e-prints, \href
  {https://ui.adsabs.harvard.edu/abs/2020arXiv201001509B} {p. arXiv:2010.01509}

\bibitem[\protect\citeauthoryear{{Bonanno} \& {Sedrakian}}{{Bonanno} \&
  {Sedrakian}}{2012}]{Bonanno2012A&A}
{Bonanno} L.,  {Sedrakian} A.,  2012, \mn@doi [\aap]
  {10.1051/0004-6361/201117832}, \href
  {https://ui.adsabs.harvard.edu/abs/2012A&A...539A..16B} {539, A16}

\bibitem[\protect\citeauthoryear{Cai, Fattoyev, Li  \& Newton}{Cai
  et~al.}{2015}]{Cai_PRC_2015}
Cai B.-J.,  Fattoyev F.~J.,  Li B.-A.,   Newton W.~G.,  2015, \mn@doi [Phys.
  Rev. C] {10.1103/PhysRevC.92.015802}, 92, 015802

\bibitem[\protect\citeauthoryear{{Carlson}, {Gandolfi}, {Pederiva}, {Pieper},
  {Schiavilla}, {Schmidt}  \& {Wiringa}}{{Carlson}
  et~al.}{2015}]{2015RvMP...87.1067C}
{Carlson} J.,  {Gandolfi} S.,  {Pederiva} F.,  {Pieper} S.~C.,  {Schiavilla}
  R.,  {Schmidt} K.~E.,   {Wiringa} R.~B.,  2015, \mn@doi [Reviews of Modern
  Physics] {10.1103/RevModPhys.87.1067}, \href
  {https://ui.adsabs.harvard.edu/abs/2015RvMP...87.1067C} {87, 1067}

\bibitem[\protect\citeauthoryear{{Chatterjee} \& {Vida{\~n}a}}{{Chatterjee} \&
  {Vida{\~n}a}}{2016}]{2016EPJA...52...29C}
{Chatterjee} D.,  {Vida{\~n}a} I.,  2016, \mn@doi [European Physical Journal A]
  {10.1140/epja/i2016-16029-x}, \href
  {https://ui.adsabs.harvard.edu/abs/2016EPJA...52...29C} {52, 29}

\bibitem[\protect\citeauthoryear{Chen, Guo  \& Liu}{Chen
  et~al.}{2007}]{PhysRevC.75.035806}
Chen Y.,  Guo H.,   Liu Y.,  2007, \mn@doi [Phys. Rev. C]
  {10.1103/PhysRevC.75.035806}, 75, 035806

\bibitem[\protect\citeauthoryear{Colucci \& Sedrakian}{Colucci \&
  Sedrakian}{2013}]{Colucci_PRC_2013}
Colucci G.,  Sedrakian A.,  2013, \mn@doi [Phys. Rev. C]
  {10.1103/PhysRevC.87.055806}, 87, 055806

\bibitem[\protect\citeauthoryear{{Cozma} \& {Tsang}}{{Cozma} \&
  {Tsang}}{2021}]{2021arXiv210108679C}
{Cozma} M.~D.,  {Tsang} M.~B.,  2021, arXiv e-prints, \href
  {https://ui.adsabs.harvard.edu/abs/2021arXiv210108679C} {p. arXiv:2101.08679}

\bibitem[\protect\citeauthoryear{{Cromartie} et~al.,}{{Cromartie}
  et~al.}{2020}]{2020NatAs...4...72C}
{Cromartie} H.~T.,  et~al., 2020, \mn@doi [Nature Astronomy]
  {10.1038/s41550-019-0880-2}, \href
  {https://ui.adsabs.harvard.edu/abs/2020NatAs...4...72C} {4, 72}

\bibitem[\protect\citeauthoryear{Damour \& Nagar}{Damour \&
  Nagar}{2010}]{PhysRevD.81.084016}
Damour T.,  Nagar A.,  2010, \mn@doi [Phys. Rev. D]
  {10.1103/PhysRevD.81.084016}, 81, 084016

\bibitem[\protect\citeauthoryear{{Demorest}, {Pennucci}, {Ransom}, {Roberts}
  \& {Hessels}}{{Demorest} et~al.}{2010}]{2010Natur.467.1081D}
{Demorest} P.~B.,  {Pennucci} T.,  {Ransom} S.~M.,  {Roberts} M.~S.~E.,
  {Hessels} J.~W.~T.,  2010, \mn@doi [\nat] {10.1038/nature09466}, \href
  {https://ui.adsabs.harvard.edu/abs/2010Natur.467.1081D} {467, 1081}

\bibitem[\protect\citeauthoryear{{Dexheimer}, {Gomes}, {Kl{\"a}hn}, {Han}  \&
  {Salinas}}{{Dexheimer} et~al.}{2021}]{2021PhRvC.103b5808D}
{Dexheimer} V.,  {Gomes} R.~O.,  {Kl{\"a}hn} T.,  {Han} S.,   {Salinas} M.,
  2021, \mn@doi [\prc] {10.1103/PhysRevC.103.025808}, \href
  {https://ui.adsabs.harvard.edu/abs/2021PhRvC.103b5808D} {103, 025808}

\bibitem[\protect\citeauthoryear{{Douchin} \& {Haensel}}{{Douchin} \&
  {Haensel}}{2001}]{2001A&A...380..151D}
{Douchin} F.,  {Haensel} P.,  2001, \mn@doi [\aap]
  {10.1051/0004-6361:20011402}, \href
  {https://ui.adsabs.harvard.edu/abs/2001A&A...380..151D} {380, 151}

\bibitem[\protect\citeauthoryear{Drago, Lavagno, Pagliara  \& Pigato}{Drago
  et~al.}{2014}]{Drago_PRC_2014}
Drago A.,  Lavagno A.,  Pagliara G.,   Pigato D.,  2014, \mn@doi [Phys. Rev. C]
  {10.1103/PhysRevC.90.065809}, 90, 065809

\bibitem[\protect\citeauthoryear{{Fattoyev}, {Horowitz}, {Piekarewicz}  \&
  {Reed}}{{Fattoyev} et~al.}{2020}]{2020PhRvC.102f5805F}
{Fattoyev} F.~J.,  {Horowitz} C.~J.,  {Piekarewicz} J.,   {Reed} B.,  2020,
  \mn@doi [\prc] {10.1103/PhysRevC.102.065805}, \href
  {https://ui.adsabs.harvard.edu/abs/2020PhRvC.102f5805F} {102, 065805}

\bibitem[\protect\citeauthoryear{Favata}{Favata}{2014}]{PhysRevLett.112.101101}
Favata M.,  2014, \mn@doi [Phys. Rev. Lett.] {10.1103/PhysRevLett.112.101101},
  112, 101101

\bibitem[\protect\citeauthoryear{Feliciello \& Nagae}{Feliciello \&
  Nagae}{2015}]{Feliciello_2015}
Feliciello A.,  Nagae T.,  2015, \mn@doi [Reports on Progress in Physics]
  {10.1088/0034-4885/78/9/096301}, 78, 096301

\bibitem[\protect\citeauthoryear{Flanagan \& Hinderer}{Flanagan \&
  Hinderer}{2008}]{PhysRevD.77.021502}
Flanagan E.~E.,  Hinderer T.,  2008, \mn@doi [Phys. Rev. D]
  {10.1103/PhysRevD.77.021502}, 77, 021502

\bibitem[\protect\citeauthoryear{{Fonseca} et~al.,}{{Fonseca}
  et~al.}{2021}]{2021arXiv210400880F}
{Fonseca} E.,  et~al., 2021, arXiv e-prints, \href
  {https://ui.adsabs.harvard.edu/abs/2021arXiv210400880F} {p. arXiv:2104.00880}

\bibitem[\protect\citeauthoryear{{Fortin}, {Provid{\^e}ncia}, {Raduta},
  {Gulminelli}, {Zdunik}, {Haensel}  \& {Bejger}}{{Fortin}
  et~al.}{2016}]{2016PhRvC..94c5804F}
{Fortin} M.,  {Provid{\^e}ncia} C.,  {Raduta} A.~R.,  {Gulminelli} F.,
  {Zdunik} J.~L.,  {Haensel} P.,   {Bejger} M.,  2016, \mn@doi [\prc]
  {10.1103/PhysRevC.94.035804}, \href
  {https://ui.adsabs.harvard.edu/abs/2016PhRvC..94c5804F} {94, 035804}

\bibitem[\protect\citeauthoryear{{Friedman} \& {Gal}}{{Friedman} \&
  {Gal}}{2021}]{2021arXiv210400421F}
{Friedman} E.,  {Gal} A.,  2021, arXiv e-prints, \href
  {https://ui.adsabs.harvard.edu/abs/2021arXiv210400421F} {p. arXiv:2104.00421}

\bibitem[\protect\citeauthoryear{Gal, Hungerford  \& Millener}{Gal
  et~al.}{2016}]{RevModPhys.88.035004}
Gal A.,  Hungerford E.~V.,   Millener D.~J.,  2016, \mn@doi [Rev. Mod. Phys.]
  {10.1103/RevModPhys.88.035004}, 88, 035004

\bibitem[\protect\citeauthoryear{{Glendenning}}{{Glendenning}}{1996}]{1996cost.book.....G}
{Glendenning} N.~K.,  1996, {Compact Stars, (Springer-Verlag, New York, 2007),
  2nd ed.}

\bibitem[\protect\citeauthoryear{Glendenning \& Moszkowski}{Glendenning \&
  Moszkowski}{1991}]{PhysRevLett.67.2414}
Glendenning N.~K.,  Moszkowski S.~A.,  1991, \mn@doi [Phys. Rev. Lett.]
  {10.1103/PhysRevLett.67.2414}, 67, 2414

\bibitem[\protect\citeauthoryear{{Glendenning} \&
  {Schaffner-Bielich}}{{Glendenning} \&
  {Schaffner-Bielich}}{1999}]{1999PhRvC..60b5803G}
{Glendenning} N.~K.,  {Schaffner-Bielich} J.,  1999, \mn@doi [\prc]
  {10.1103/PhysRevC.60.025803}, \href
  {https://ui.adsabs.harvard.edu/abs/1999PhRvC..60b5803G} {60, 025803}

\bibitem[\protect\citeauthoryear{{Gomes}, {Dexheimer}, {Schramm}  \&
  {Vasconcellos}}{{Gomes} et~al.}{2015}]{2015ApJ...808....8G}
{Gomes} R.~O.,  {Dexheimer} V.,  {Schramm} S.,   {Vasconcellos} C.~A.~Z.,
  2015, \mn@doi [\apj] {10.1088/0004-637X/808/1/8}, \href
  {https://ui.adsabs.harvard.edu/abs/2015ApJ...808....8G} {808, 8}

\bibitem[\protect\citeauthoryear{{Gomes}, {Char}  \& {Schramm}}{{Gomes}
  et~al.}{2019}]{2019ApJ...877..139G}
{Gomes} R.~O.,  {Char} P.,   {Schramm} S.,  2019, \mn@doi [\apj]
  {10.3847/1538-4357/ab1751}, \href
  {https://ui.adsabs.harvard.edu/abs/2019ApJ...877..139G} {877, 139}

\bibitem[\protect\citeauthoryear{{Haensel} \& {Proszynski}}{{Haensel} \&
  {Proszynski}}{1982}]{1982ApJ...258..306H}
{Haensel} P.,  {Proszynski} M.,  1982, \mn@doi [\apj] {10.1086/160080}, \href
  {https://ui.adsabs.harvard.edu/abs/1982ApJ...258..306H} {258, 306}

\bibitem[\protect\citeauthoryear{Hinderer}{Hinderer}{2008}]{Hinderer_2008}
Hinderer T.,  2008, \mn@doi [The Astrophysical Journal] {10.1086/533487}, 677,
  1216

\bibitem[\protect\citeauthoryear{Hinderer, Lackey, Lang  \& Read}{Hinderer
  et~al.}{2010}]{PhysRevD.81.123016}
Hinderer T.,  Lackey B.~D.,  Lang R.~N.,   Read J.~S.,  2010, \mn@doi [Phys.
  Rev. D] {10.1103/PhysRevD.81.123016}, 81, 123016

\bibitem[\protect\citeauthoryear{{Hofmann}, {Keil}  \& {Lenske}}{{Hofmann}
  et~al.}{2001}]{2001PhRvC..64b5804H}
{Hofmann} F.,  {Keil} C.~M.,   {Lenske} H.,  2001, \mn@doi [\prc]
  {10.1103/PhysRevC.64.025804}, \href
  {https://ui.adsabs.harvard.edu/abs/2001PhRvC..64b5804H} {64, 025804}

\bibitem[\protect\citeauthoryear{Jiang, Tang, Wang, Fan  \& Wei}{Jiang
  et~al.}{2020}]{Jiang_2020}
Jiang J.-L.,  Tang S.-P.,  Wang Y.-Z.,  Fan Y.-Z.,   Wei D.-M.,  2020, \mn@doi
  [The Astrophysical Journal] {10.3847/1538-4357/ab77cf}, 892, 55

\bibitem[\protect\citeauthoryear{Kanakis-Pegios \& Moustakidis}{Kanakis-Pegios
  \& Moustakidis}{2020}]{hnps2989}
Kanakis-Pegios A.,  Moustakidis C.,  2020, \mn@doi [HNPS Advances in Nuclear
  Physics] {10.12681/hnps.2989}, 27, 95

\bibitem[\protect\citeauthoryear{Kl\"ahn et~al.,}{Kl\"ahn
  et~al.}{2006}]{PhysRevC.74.035802}
Kl\"ahn T.,  et~al., 2006, \mn@doi [Phys. Rev. C] {10.1103/PhysRevC.74.035802},
  74, 035802

\bibitem[\protect\citeauthoryear{Koch \& Ohtsuka}{Koch \&
  Ohtsuka}{1985}]{KOCH1985765}
Koch J.,  Ohtsuka N.,  1985, \mn@doi [Nuclear Physics A]
  {https://doi.org/10.1016/0375-9474(85)90187-3}, 435, 765

\bibitem[\protect\citeauthoryear{Kolomeitsev, Maslov  \&
  Voskresensky}{Kolomeitsev et~al.}{2017}]{Kolomeitsev_NPA_2017}
Kolomeitsev E.,  Maslov K.,   Voskresensky D.,  2017, \mn@doi [Nuclear Physics
  A] {https://doi.org/10.1016/j.nuclphysa.2017.02.004}, 961, 106

\bibitem[\protect\citeauthoryear{Kumar, Biswal  \& Patra}{Kumar
  et~al.}{2017}]{PhysRevC.95.015801}
Kumar B.,  Biswal S.~K.,   Patra S.~K.,  2017, \mn@doi [Phys. Rev. C]
  {10.1103/PhysRevC.95.015801}, 95, 015801

\bibitem[\protect\citeauthoryear{{Lalazissis}, {K{\"o}nig}  \&
  {Ring}}{{Lalazissis} et~al.}{1997}]{1997PhRvC..55..540L}
{Lalazissis} G.~A.,  {K{\"o}nig} J.,   {Ring} P.,  1997, \mn@doi [\prc]
  {10.1103/PhysRevC.55.540}, \href
  {https://ui.adsabs.harvard.edu/abs/1997PhRvC..55..540L} {55, 540}

\bibitem[\protect\citeauthoryear{{Lalazissis}, {Nik{\v{s}}i{\'c}}, {Vretenar}
  \& {Ring}}{{Lalazissis} et~al.}{2005}]{2005PhRvC..71b4312L}
{Lalazissis} G.~A.,  {Nik{\v{s}}i{\'c}} T.,  {Vretenar} D.,   {Ring} P.,  2005,
  \mn@doi [\prc] {10.1103/PhysRevC.71.024312}, \href
  {https://ui.adsabs.harvard.edu/abs/2005PhRvC..71b4312L} {71, 024312}

\bibitem[\protect\citeauthoryear{Lalazissis, Karatzikos, Fossion, Arteaga,
  Afanasjev  \& Ring}{Lalazissis et~al.}{2009}]{LALAZISSIS200936}
Lalazissis G.,  Karatzikos S.,  Fossion R.,  Arteaga D.~P.,  Afanasjev A.,
  Ring P.,  2009, \mn@doi [Physics Letters B]
  {https://doi.org/10.1016/j.physletb.2008.11.070}, 671, 36

\bibitem[\protect\citeauthoryear{{Landry}, {Essick}  \&
  {Chatziioannou}}{{Landry} et~al.}{2020}]{2020PhRvD.101l3007L}
{Landry} P.,  {Essick} R.,   {Chatziioannou} K.,  2020, \mn@doi [\prd]
  {10.1103/PhysRevD.101.123007}, \href
  {https://ui.adsabs.harvard.edu/abs/2020PhRvD.101l3007L} {101, 123007}

\bibitem[\protect\citeauthoryear{{Lattimer} \& {Prakash}}{{Lattimer} \&
  {Prakash}}{2016}]{2016PhR...621..127L}
{Lattimer} J.~M.,  {Prakash} M.,  2016, \mn@doi [\physrep]
  {10.1016/j.physrep.2015.12.005}, \href
  {https://ui.adsabs.harvard.edu/abs/2016PhR...621..127L} {621, 127}

\bibitem[\protect\citeauthoryear{{Li} \& {Sedrakian}}{{Li} \&
  {Sedrakian}}{2019}]{Li2019ApJ}
{Li} J.~J.,  {Sedrakian} A.,  2019, \mn@doi [\apjl] {10.3847/2041-8213/ab1090},
  \href {https://ui.adsabs.harvard.edu/abs/2019ApJ...874L..22L} {874, L22}

\bibitem[\protect\citeauthoryear{{Li} \& {Steiner}}{{Li} \&
  {Steiner}}{2006}]{2006PhLB..642..436L}
{Li} B.-A.,  {Steiner} A.~W.,  2006, \mn@doi [Physics Letters B]
  {10.1016/j.physletb.2006.09.065}, \href
  {https://ui.adsabs.harvard.edu/abs/2006PhLB..642..436L} {642, 436}

\bibitem[\protect\citeauthoryear{{Li}, {Long}  \& {Sedrakian}}{{Li}
  et~al.}{2018a}]{2018EPJA...54..133L}
{Li} J.~J.,  {Long} W.~H.,   {Sedrakian} A.,  2018a, \mn@doi [European Physical
  Journal A] {10.1140/epja/i2018-12566-6}, \href
  {https://ui.adsabs.harvard.edu/abs/2018EPJA...54..133L} {54, 133}

\bibitem[\protect\citeauthoryear{Li, Sedrakian  \& Weber}{Li
  et~al.}{2018b}]{Li_PLB_2018}
Li J.~J.,  Sedrakian A.,   Weber F.,  2018b, \mn@doi [Phys. Lett. B]
  {https://doi.org/10.1016/j.physletb.2018.06.051}, 783, 234

\bibitem[\protect\citeauthoryear{{Li}, {Sedrakian}  \& {Alford}}{{Li}
  et~al.}{2020a}]{Li2020PhRvD}
{Li} J.~J.,  {Sedrakian} A.,   {Alford} M.,  2020a, \mn@doi [\prd]
  {10.1103/PhysRevD.101.063022}, \href
  {https://ui.adsabs.harvard.edu/abs/2020PhRvD.101f3022L} {101, 063022}

\bibitem[\protect\citeauthoryear{Li, Sedrakian  \& Weber}{Li
  et~al.}{2020b}]{LI2020135812}
Li J.~J.,  Sedrakian A.,   Weber F.,  2020b, \mn@doi [Physics Letters B]
  {https://doi.org/10.1016/j.physletb.2020.135812}, 810, 135812

\bibitem[\protect\citeauthoryear{{Li}, {Cai}, {Xie}  \& {Zhang}}{{Li}
  et~al.}{2021a}]{2021arXiv210504629L}
{Li} B.-A.,  {Cai} B.-J.,  {Xie} W.-J.,   {Zhang} N.-B.,  2021a, arXiv
  e-prints, \href {https://ui.adsabs.harvard.edu/abs/2021arXiv210504629L} {p.
  arXiv:2105.04629}

\bibitem[\protect\citeauthoryear{{Li}, {Chen}, {Wen}  \& {Zhang}}{{Li}
  et~al.}{2021b}]{2021EPJA...57...31L}
{Li} Y.,  {Chen} H.,  {Wen} D.,   {Zhang} J.,  2021b, \mn@doi [European
  Physical Journal A] {10.1140/epja/s10050-021-00342-w}, \href
  {https://ui.adsabs.harvard.edu/abs/2021EPJA...57...31L} {57, 31}

\bibitem[\protect\citeauthoryear{{Logoteta}}{{Logoteta}}{2019}]{2019PhRvC.100d5803L}
{Logoteta} D.,  2019, \mn@doi [\prc] {10.1103/PhysRevC.100.045803}, \href
  {https://ui.adsabs.harvard.edu/abs/2019PhRvC.100d5803L} {100, 045803}

\bibitem[\protect\citeauthoryear{{Long}, {Meng}, {Giai}  \& {Zhou}}{{Long}
  et~al.}{2004}]{2004PhRvC..69c4319L}
{Long} W.,  {Meng} J.,  {Giai} N.~V.,   {Zhou} S.-G.,  2004, \mn@doi [\prc]
  {10.1103/PhysRevC.69.034319}, \href
  {https://ui.adsabs.harvard.edu/abs/2004PhRvC..69c4319L} {69, 034319}

\bibitem[\protect\citeauthoryear{Malik, Alam, Fortin, Provid\^encia, Agrawal,
  Jha, Kumar  \& Patra}{Malik et~al.}{2018}]{PhysRevC.98.035804}
Malik T.,  Alam N.,  Fortin M.,  Provid\^encia C.,  Agrawal B.~K.,  Jha T.~K.,
  Kumar B.,   Patra S.~K.,  2018, \mn@doi [Phys. Rev. C]
  {10.1103/PhysRevC.98.035804}, 98, 035804

\bibitem[\protect\citeauthoryear{{Malik}, {Banik}  \& {Bandyopadhyay}}{{Malik}
  et~al.}{2021}]{2021EPJST.tmp...32M}
{Malik} T.,  {Banik} S.,   {Bandyopadhyay} D.,  2021, \mn@doi [European
  Physical Journal Special Topics] {10.1140/epjs/s11734-021-00006-2}, \href
  {https://ui.adsabs.harvard.edu/abs/2021EPJST.tmp...32M} {}

\bibitem[\protect\citeauthoryear{Mannarelli}{Mannarelli}{2019}]{particles2030025}
Mannarelli M.,  2019, \mn@doi [Particles] {10.3390/particles2030025}, 2, 411

\bibitem[\protect\citeauthoryear{{Mariani}, {Orsaria}, {Ranea-Sandoval}  \&
  {Lugones}}{{Mariani} et~al.}{2019}]{2019MNRAS.489.4261M}
{Mariani} M.,  {Orsaria} M.~G.,  {Ranea-Sandoval} I.~F.,   {Lugones} G.,  2019,
  \mn@doi [\mnras] {10.1093/mnras/stz2392}, \href
  {https://ui.adsabs.harvard.edu/abs/2019MNRAS.489.4261M} {489, 4261}

\bibitem[\protect\citeauthoryear{{Maselli}, {Cardoso}, {Ferrari}, {Gualtieri}
  \& {Pani}}{{Maselli} et~al.}{2013}]{2013PhRvD..88b3007M}
{Maselli} A.,  {Cardoso} V.,  {Ferrari} V.,  {Gualtieri} L.,   {Pani} P.,
  2013, \mn@doi [\prd] {10.1103/PhysRevD.88.023007}, \href
  {https://ui.adsabs.harvard.edu/abs/2013PhRvD..88b3007M} {88, 023007}

\bibitem[\protect\citeauthoryear{{Miller} et~al.,}{{Miller}
  et~al.}{2019}]{2019ApJ...887L..24M}
{Miller} M.~C.,  et~al., 2019, \mn@doi [\apjl] {10.3847/2041-8213/ab50c5},
  \href {https://ui.adsabs.harvard.edu/abs/2019ApJ...887L..24M} {887, L24}

\bibitem[\protect\citeauthoryear{{Miller} et~al.,}{{Miller}
  et~al.}{2021}]{2021arXiv210506979M}
{Miller} M.~C.,  et~al., 2021, arXiv e-prints, \href
  {https://ui.adsabs.harvard.edu/abs/2021arXiv210506979M} {p. arXiv:2105.06979}

\bibitem[\protect\citeauthoryear{{Mondal}, {Agrawal}, {De}, {Samaddar},
  {Centelles}  \& {Vi{\~n}as}}{{Mondal} et~al.}{2017}]{2017PhRvC..96b1302M}
{Mondal} C.,  {Agrawal} B.~K.,  {De} J.~N.,  {Samaddar} S.~K.,  {Centelles} M.,
    {Vi{\~n}as} X.,  2017, \mn@doi [\prc] {10.1103/PhysRevC.96.021302}, \href
  {https://ui.adsabs.harvard.edu/abs/2017PhRvC..96b1302M} {96, 021302}

\bibitem[\protect\citeauthoryear{{Most}, {Papenfort}, {Weih}  \&
  {Rezzolla}}{{Most} et~al.}{2020}]{2020MNRAS.499L..82M}
{Most} E.~R.,  {Papenfort} L.~J.,  {Weih} L.~R.,   {Rezzolla} L.,  2020,
  \mn@doi [\mnras] {10.1093/mnrasl/slaa168}, \href
  {https://ui.adsabs.harvard.edu/abs/2020MNRAS.499L..82M} {499, L82}

\bibitem[\protect\citeauthoryear{{Motta}, {Thomas}  \& {Guichon}}{{Motta}
  et~al.}{2020}]{2020PhLB..80235266M}
{Motta} T.~F.,  {Thomas} A.~W.,   {Guichon} P.~A.~M.,  2020, \mn@doi [Physics
  Letters B] {10.1016/j.physletb.2020.135266}, \href
  {https://ui.adsabs.harvard.edu/abs/2020PhLB..80235266M} {802, 135266}

\bibitem[\protect\citeauthoryear{Nakamura, Sato, Lee, Szczerbinska  \&
  Kubodera}{Nakamura et~al.}{2010}]{PhysRevC.81.035502}
Nakamura S.~X.,  Sato T.,  Lee T.-S.~H.,  Szczerbinska B.,   Kubodera K.,
  2010, \mn@doi [Phys. Rev. C] {10.1103/PhysRevC.81.035502}, 81, 035502

\bibitem[\protect\citeauthoryear{Nandi \& Char}{Nandi \&
  Char}{2018}]{Nandi_2018}
Nandi R.,  Char P.,  2018, \mn@doi [The Astrophysical Journal]
  {10.3847/1538-4357/aab78c}, 857, 12

\bibitem[\protect\citeauthoryear{{Nandi} \& {Pal}}{{Nandi} \&
  {Pal}}{2021}]{2021EPJST.tmp....5N}
{Nandi} R.,  {Pal} S.,  2021, \mn@doi [European Physical Journal Special
  Topics] {10.1140/epjs/s11734-021-00004-4}, \href
  {https://ui.adsabs.harvard.edu/abs/2021EPJST.tmp....5N} {}

\bibitem[\protect\citeauthoryear{Nandi, Char  \& Pal}{Nandi
  et~al.}{2019}]{PhysRevC.99.052802}
Nandi R.,  Char P.,   Pal S.,  2019, \mn@doi [Phys. Rev. C]
  {10.1103/PhysRevC.99.052802}, 99, 052802

\bibitem[\protect\citeauthoryear{Nik\ifmmode \check{s}\else
  \v{s}\fi{}i\ifmmode~\acute{c}\else \'{c}\fi{}, Vretenar, Finelli  \&
  Ring}{Nik\ifmmode \check{s}\else \v{s}\fi{}i\ifmmode~\acute{c}\else
  \'{c}\fi{} et~al.}{2002}]{PhysRevC.66.024306}
Nik\ifmmode \check{s}\else \v{s}\fi{}i\ifmmode~\acute{c}\else \'{c}\fi{} T.,
  Vretenar D.,  Finelli P.,   Ring P.,  2002, \mn@doi [Phys. Rev. C]
  {10.1103/PhysRevC.66.024306}, 66, 024306

\bibitem[\protect\citeauthoryear{Oertel, Provid{\^e}ncia, Gulminelli  \&
  Raduta}{Oertel et~al.}{2015}]{Oertel2015}
Oertel M.,  Provid{\^e}ncia C.,  Gulminelli F.,   Raduta A.~R.,  2015, J. Phys.
  G, 42, 075202

\bibitem[\protect\citeauthoryear{{Oertel}, {Hempel}, {Kl{\"a}hn}  \&
  {Typel}}{{Oertel} et~al.}{2017}]{2017RvMP...89a5007O}
{Oertel} M.,  {Hempel} M.,  {Kl{\"a}hn} T.,   {Typel} S.,  2017, \mn@doi
  [Reviews of Modern Physics] {10.1103/RevModPhys.89.015007}, \href
  {https://ui.adsabs.harvard.edu/abs/2017RvMP...89a5007O} {89, 015007}

\bibitem[\protect\citeauthoryear{{Pal}, {Bandyopadhyay}  \& {Greiner}}{{Pal}
  et~al.}{2000}]{2000NuPhA.674..553P}
{Pal} S.,  {Bandyopadhyay} D.,   {Greiner} W.,  2000, \mn@doi [\nphysa]
  {10.1016/S0375-9474(00)00175-5}, \href
  {https://ui.adsabs.harvard.edu/abs/2000NuPhA.674..553P} {674, 553}

\bibitem[\protect\citeauthoryear{{Pang}, {Tews}, {Coughlin}, {Bulla}, {Van Den
  Broeck}  \& {Dietrich}}{{Pang} et~al.}{2021}]{2021arXiv210508688P}
{Pang} P. T.~H.,  {Tews} I.,  {Coughlin} M.~W.,  {Bulla} M.,  {Van Den Broeck}
  C.,   {Dietrich} T.,  2021, arXiv e-prints, \href
  {https://ui.adsabs.harvard.edu/abs/2021arXiv210508688P} {p. arXiv:2105.08688}

\bibitem[\protect\citeauthoryear{{Paschalidis}, {Yagi}, {Alvarez-Castillo},
  {Blaschke}  \& {Sedrakian}}{{Paschalidis} et~al.}{2018}]{2018PhRvD..97h4038P}
{Paschalidis} V.,  {Yagi} K.,  {Alvarez-Castillo} D.,  {Blaschke} D.~B.,
  {Sedrakian} A.,  2018, \mn@doi [\prd] {10.1103/PhysRevD.97.084038}, \href
  {https://ui.adsabs.harvard.edu/abs/2018PhRvD..97h4038P} {97, 084038}

\bibitem[\protect\citeauthoryear{{Prakash}, {Bombaci}, {Prakash}, {Ellis},
  {Lattimer}  \& {Knorren}}{{Prakash} et~al.}{1997}]{1997PhR...280....1P}
{Prakash} M.,  {Bombaci} I.,  {Prakash} M.,  {Ellis} P.~J.,  {Lattimer} J.~M.,
   {Knorren} R.,  1997, \mn@doi [\physrep] {10.1016/S0370-1573(96)00023-3},
  \href {https://ui.adsabs.harvard.edu/abs/1997PhR...280....1P} {280, 1}

\bibitem[\protect\citeauthoryear{{Raaijmakers} et~al.,}{{Raaijmakers}
  et~al.}{2021}]{2021arXiv210506981R}
{Raaijmakers} G.,  et~al., 2021, arXiv e-prints, \href
  {https://ui.adsabs.harvard.edu/abs/2021arXiv210506981R} {p. arXiv:2105.06981}

\bibitem[\protect\citeauthoryear{Radice, Perego, Zappa  \& Bernuzzi}{Radice
  et~al.}{2018}]{Radice2018}
Radice D.,  Perego A.,  Zappa F.,   Bernuzzi S.,  2018, ApJL, 852, L29

\bibitem[\protect\citeauthoryear{{Raduta}}{{Raduta}}{2021}]{2021PhLB..81436070R}
{Raduta} A.~R.,  2021, \mn@doi [Physics Letters B]
  {10.1016/j.physletb.2021.136070}, \href
  {https://ui.adsabs.harvard.edu/abs/2021PhLB..81436070R} {814, 136070}

\bibitem[\protect\citeauthoryear{{Raduta}, {Sedrakian}  \& {Weber}}{{Raduta}
  et~al.}{2018}]{2018MNRAS.475.4347R}
{Raduta} A.~R.,  {Sedrakian} A.,   {Weber} F.,  2018, \mn@doi [\mnras]
  {10.1093/mnras/stx3318}, \href
  {https://ui.adsabs.harvard.edu/abs/2018MNRAS.475.4347R} {475, 4347}

\bibitem[\protect\citeauthoryear{Raithel, Özel  \& Psaltis}{Raithel
  et~al.}{2018}]{Raithel_2018}
Raithel C.~A.,  Özel F.,   Psaltis D.,  2018, \mn@doi [The Astrophysical
  Journal] {10.3847/2041-8213/aabcbf}, 857, L23

\bibitem[\protect\citeauthoryear{Rashdan}{Rashdan}{2001}]{PhysRevC.63.044303}
Rashdan M.,  2001, \mn@doi [Phys. Rev. C] {10.1103/PhysRevC.63.044303}, 63,
  044303

\bibitem[\protect\citeauthoryear{Rather, Rahaman, Imran, Das, Usmani  \&
  Patra}{Rather et~al.}{2021}]{PhysRevC.103.055814}
Rather I.~A.,  Rahaman U.,  Imran M.,  Das H.~C.,  Usmani A.~A.,   Patra S.~K.,
   2021, \mn@doi [Phys. Rev. C] {10.1103/PhysRevC.103.055814}, 103, 055814

\bibitem[\protect\citeauthoryear{Ribes, Ramos, Tolos, Gonzalez-Boquera  \&
  Centelles}{Ribes et~al.}{2019}]{Ribes_2019}
Ribes P.,  Ramos A.,  Tolos L.,  Gonzalez-Boquera C.,   Centelles M.,  2019,
  \mn@doi [ApJ] {10.3847/1538-4357/ab3a93}, 883, 168

\bibitem[\protect\citeauthoryear{{Riley} et~al.,}{{Riley}
  et~al.}{2019}]{2019ApJ...887L..21R}
{Riley} T.~E.,  et~al., 2019, \mn@doi [\apjl] {10.3847/2041-8213/ab481c}, \href
  {https://ui.adsabs.harvard.edu/abs/2019ApJ...887L..21R} {887, L21}

\bibitem[\protect\citeauthoryear{{Riley} et~al.,}{{Riley}
  et~al.}{2021}]{2021arXiv210506980R}
{Riley} T.~E.,  et~al., 2021, arXiv e-prints, \href
  {https://ui.adsabs.harvard.edu/abs/2021arXiv210506980R} {p. arXiv:2105.06980}

\bibitem[\protect\citeauthoryear{{Romani}, {Kandel}, {Filippenko}, {Brink}  \&
  {Zheng}}{{Romani} et~al.}{2021}]{2021ApJ...908L..46R}
{Romani} R.~W.,  {Kandel} D.,  {Filippenko} A.~V.,  {Brink} T.~G.,   {Zheng}
  W.,  2021, \mn@doi [\apjl] {10.3847/2041-8213/abe2b4}, \href
  {https://ui.adsabs.harvard.edu/abs/2021ApJ...908L..46R} {908, L46}

\bibitem[\protect\citeauthoryear{Sahoo, Mitra, Mishra, Panda  \& Li}{Sahoo
  et~al.}{2018}]{Sahoo_PRC_2018}
Sahoo H.~S.,  Mitra G.,  Mishra R.,  Panda P.~K.,   Li B.-A.,  2018, \mn@doi
  [Phys. Rev. C] {10.1103/PhysRevC.98.045801}, 98, 045801

\bibitem[\protect\citeauthoryear{{Schaffner-Bielich} \&
  {Gal}}{{Schaffner-Bielich} \& {Gal}}{2000}]{2000PhRvC..62c4311S}
{Schaffner-Bielich} J.,  {Gal} A.,  2000, \mn@doi [\prc]
  {10.1103/PhysRevC.62.034311}, \href
  {https://ui.adsabs.harvard.edu/abs/2000PhRvC..62c4311S} {62, 034311}

\bibitem[\protect\citeauthoryear{Schaffner, Dover, Gal, Greiner, Millener  \&
  Stocker}{Schaffner et~al.}{1994}]{SCHAFFNER199435}
Schaffner J.,  Dover C.,  Gal A.,  Greiner C.,  Millener D.,   Stocker H.,
  1994, \mn@doi [Annals of Physics] {https://doi.org/10.1006/aphy.1994.1090},
  235, 35

\bibitem[\protect\citeauthoryear{{Sedrakian}}{{Sedrakian}}{2007}]{Sedrakian2007PrPNP}
{Sedrakian} A.,  2007, \mn@doi [Progress in Particle and Nuclear Physics]
  {10.1016/j.ppnp.2006.02.002}, \href
  {https://ui.adsabs.harvard.edu/abs/2007PrPNP..58..168S} {58, 168}

\bibitem[\protect\citeauthoryear{{Sedrakian, Armen}}{{Sedrakian,
  Armen}}{2017}]{Sedrakian2017_EPJWeb}
{Sedrakian, Armen} 2017, \mn@doi [EPJ Web Conf.]
  {10.1051/epjconf/201716401009}, 164, 01009

\bibitem[\protect\citeauthoryear{{Sedrakian}, {Weber}  \& {Li}}{{Sedrakian}
  et~al.}{2020}]{2020PhRvD.102d1301S}
{Sedrakian} A.,  {Weber} F.,   {Li} J.~J.,  2020, \mn@doi [\prd]
  {10.1103/PhysRevD.102.041301}, \href
  {https://ui.adsabs.harvard.edu/abs/2020PhRvD.102d1301S} {102, 041301}

\bibitem[\protect\citeauthoryear{{Sedrakian}, {Li}  \& {Weber}}{{Sedrakian}
  et~al.}{2021}]{2021arXiv210514050S}
{Sedrakian} A.,  {Li} J.-J.,   {Weber} F.,  2021, arXiv e-prints, \href
  {https://ui.adsabs.harvard.edu/abs/2021arXiv210514050S} {p. arXiv:2105.14050}

\bibitem[\protect\citeauthoryear{{Serot} \& {Walecka}}{{Serot} \&
  {Walecka}}{1997}]{1997IJMPE...6..515S}
{Serot} B.~D.,  {Walecka} J.~D.,  1997, \mn@doi [International Journal of
  Modern Physics E] {10.1142/S0218301397000299}, \href
  {https://ui.adsabs.harvard.edu/abs/1997IJMPE...6..515S} {6, 515}

\bibitem[\protect\citeauthoryear{Sharma, Nagarajan  \& Ring}{Sharma
  et~al.}{1993}]{SHARMA1993377}
Sharma M.,  Nagarajan M.,   Ring P.,  1993, \mn@doi [Physics Letters B]
  {https://doi.org/10.1016/0370-2693(93)90970-S}, 312, 377

\bibitem[\protect\citeauthoryear{{Somasundaram} \& {Margueron}}{{Somasundaram}
  \& {Margueron}}{2021}]{2021arXiv210413612S}
{Somasundaram} R.,  {Margueron} J.,  2021, arXiv e-prints, \href
  {https://ui.adsabs.harvard.edu/abs/2021arXiv210413612S} {p. arXiv:2104.13612}

\bibitem[\protect\citeauthoryear{Taninah, Agbemava, Afanasjev  \& Ring}{Taninah
  et~al.}{2020}]{TANINAH2020135065}
Taninah A.,  Agbemava S.,  Afanasjev A.,   Ring P.,  2020, \mn@doi [Physics
  Letters B] {https://doi.org/10.1016/j.physletb.2019.135065}, 800, 135065

\bibitem[\protect\citeauthoryear{Tews, Margueron  \& Reddy}{Tews
  et~al.}{2018}]{Tews2018}
Tews I.,  Margueron J.,   Reddy S.,  2018, PhRvC, 98, 045804

\bibitem[\protect\citeauthoryear{Tews, Pang, Dietrich, Coughlin, Antier, Bulla,
  Heinzel  \& Issa}{Tews et~al.}{2021}]{Tews_2021}
Tews I.,  Pang P. T.~H.,  Dietrich T.,  Coughlin M.~W.,  Antier S.,  Bulla M.,
  Heinzel J.,   Issa L.,  2021, \mn@doi [The Astrophysical Journal]
  {10.3847/2041-8213/abdaae}, 908, L1

\bibitem[\protect\citeauthoryear{Thapa \& Sinha}{Thapa \&
  Sinha}{2020}]{PhysRevD.102.123007}
Thapa V.~B.,  Sinha M.,  2020, \mn@doi [Phys. Rev. D]
  {10.1103/PhysRevD.102.123007}, 102, 123007

\bibitem[\protect\citeauthoryear{Thapa, Sinha, Li  \& Sedrakian}{Thapa
  et~al.}{2020}]{particles3040043}
Thapa V.~B.,  Sinha M.,  Li J.~J.,   Sedrakian A.,  2020, \mn@doi [Particles]
  {10.3390/particles3040043}, \href
  {https://ui.adsabs.harvard.edu/abs/2020arXiv201000981B} {3, 660}

\bibitem[\protect\citeauthoryear{Thapa, Sinha, Li  \& Sedrakian}{Thapa
  et~al.}{2021}]{PhysRevD.103.063004}
Thapa V.~B.,  Sinha M.,  Li J.~J.,   Sedrakian A.,  2021, \mn@doi [Phys. Rev.
  D] {10.1103/PhysRevD.103.063004}, 103, 063004

\bibitem[\protect\citeauthoryear{{Tolos} \& {Fabbietti}}{{Tolos} \&
  {Fabbietti}}{2020}]{2020PrPNP.11203770T}
{Tolos} L.,  {Fabbietti} L.,  2020, \mn@doi [Progress in Particle and Nuclear
  Physics] {10.1016/j.ppnp.2020.103770}, \href
  {https://ui.adsabs.harvard.edu/abs/2020PrPNP.11203770T} {112, 103770}

\bibitem[\protect\citeauthoryear{Tolos, Centelles  \& Ramos}{Tolos
  et~al.}{2017}]{Tolos2017b}
Tolos L.,  Centelles M.,   Ramos A.,  2017, \mn@doi [Publ. Astron. Soc.
  Austral.] {10.1017/pasa.2017.60}, 34, e065

\bibitem[\protect\citeauthoryear{Typel}{Typel}{2005}]{PhysRevC.71.064301}
Typel S.,  2005, \mn@doi [Phys. Rev. C] {10.1103/PhysRevC.71.064301}, 71,
  064301

\bibitem[\protect\citeauthoryear{Typel \& Alvear~Terrero}{Typel \&
  Alvear~Terrero}{2020}]{Typel:2020ozc}
Typel S.,  Alvear~Terrero D.,  2020, \mn@doi [Eur. Phys. J. A]
  {10.1140/epja/s10050-020-00172-2}, 56, 160

\bibitem[\protect\citeauthoryear{Typel \& Wolter}{Typel \&
  Wolter}{1999}]{TYPEL1999331}
Typel S.,  Wolter H.,  1999, \mn@doi [Nuclear Physics A]
  {https://doi.org/10.1016/S0375-9474(99)00310-3}, 656, 331

\bibitem[\protect\citeauthoryear{Typel, R\"opke, Kl\"ahn, Blaschke  \&
  Wolter}{Typel et~al.}{2010}]{PhysRevC.81.015803}
Typel S.,  R\"opke G.,  Kl\"ahn T.,  Blaschke D.,   Wolter H.~H.,  2010,
  \mn@doi [Phys. Rev. C] {10.1103/PhysRevC.81.015803}, 81, 015803

\bibitem[\protect\citeauthoryear{{Vautherin} \& {Brink}}{{Vautherin} \&
  {Brink}}{1972}]{1972PhRvC...5..626V}
{Vautherin} D.,  {Brink} D.~M.,  1972, \mn@doi [\prc] {10.1103/PhysRevC.5.626},
  \href {https://ui.adsabs.harvard.edu/abs/1972PhRvC...5..626V} {5, 626}

\bibitem[\protect\citeauthoryear{{Walecka}}{{Walecka}}{1974}]{1974AnPhy..83..491W}
{Walecka} J.~D.,  1974, \mn@doi [Annals of Physics]
  {10.1016/0003-4916(74)90208-5}, \href
  {https://ui.adsabs.harvard.edu/abs/1974AnPhy..83..491W} {83, 491}

\bibitem[\protect\citeauthoryear{Weber}{Weber}{2017}]{weber2017pulsars}
Weber F.,  2017, Pulsars as Astrophysical Laboratories for Nuclear and Particle
  Physics.
Series in High Energy Physics, Cosmology and Gravitation, CRC Press, \url
  {https://books.google.co.in/books?id=SSw2DwAAQBAJ}

\bibitem[\protect\citeauthoryear{Wehrberger, Bedau  \& Beck}{Wehrberger
  et~al.}{1989}]{WEHRBERGER1989797}
Wehrberger K.,  Bedau C.,   Beck F.,  1989, \mn@doi [Nuclear Physics A]
  {https://doi.org/10.1016/0375-9474(89)90008-0}, 504, 797

\bibitem[\protect\citeauthoryear{Weissenborn, Chatterjee  \&
  Schaffner-Bielich}{Weissenborn et~al.}{2012}]{Weissenborn2012a}
Weissenborn S.,  Chatterjee D.,   Schaffner-Bielich J.,  2012, Nucl. Phys. A,
  881, 62

\bibitem[\protect\citeauthoryear{{Yagi} \& {Yunes}}{{Yagi} \&
  {Yunes}}{2017}]{2017PhR...681....1Y}
{Yagi} K.,  {Yunes} N.,  2017, \mn@doi [\physrep]
  {10.1016/j.physrep.2017.03.002}, \href
  {https://ui.adsabs.harvard.edu/abs/2017PhR...681....1Y} {681, 1}

\bibitem[\protect\citeauthoryear{{Zhang} \& {Li}}{{Zhang} \&
  {Li}}{2021}]{2021arXiv210511031Z}
{Zhang} N.-B.,  {Li} B.-A.,  2021, arXiv e-prints, \href
  {https://ui.adsabs.harvard.edu/abs/2021arXiv210511031Z} {p. arXiv:2105.11031}

\bibitem[\protect\citeauthoryear{Zhu, Li, Hu  \& Sagawa}{Zhu
  et~al.}{2016}]{Zhu_PRC_2016}
Zhu Z.-Y.,  Li A.,  Hu J.-N.,   Sagawa H.,  2016, \mn@doi [Phys. Rev. C]
  {10.1103/PhysRevC.94.045803}, 94, 045803

\bibitem[\protect\citeauthoryear{{Zimmerman}, {Carson}, {Schumacher}, {Steiner}
   \& {Yagi}}{{Zimmerman} et~al.}{2020}]{2020arXiv200203210Z}
{Zimmerman} J.,  {Carson} Z.,  {Schumacher} K.,  {Steiner} A.~W.,   {Yagi} K.,
  2020, arXiv e-prints, \href
  {https://ui.adsabs.harvard.edu/abs/2020arXiv200203210Z} {p. arXiv:2002.03210}

\makeatother
\end{thebibliography}








\label{lastpage}
\end{document}